\documentclass[11pt]{article}
\pdfoutput=1
\usepackage{bbold}
\usepackage{amsmath}
\usepackage{url}
\usepackage{hyperref}
\usepackage{slashed}
\usepackage{tikz}
\usetikzlibrary{decorations.pathmorphing}
\usetikzlibrary{matrix}
\usepackage{graphicx}
\usepackage{epstopdf}
\usepackage{subfigure}
\usepackage{pgfplots}
\usepackage{amssymb}
\usepackage{color}
\usepackage{mathrsfs}
\usepackage{fancybox}
\usepackage{lipsum}
\usepackage{eurosym}
\usepackage{tcolorbox}
\usepackage{anyfontsize}
\usepackage{enumitem,kantlipsum}

\usetikzlibrary{shadings}

\usetikzlibrary{decorations.pathmorphing}
\tikzset{snake it/.style={decorate, decoration=snake}}
\let\originallesssim\lesssim
\let\originalgtrsim\gtrsim
\DeclareRobustCommand{\lesssim}{%
  \mathrel{\mathpalette\lowersim\originallesssim}%
}
\DeclareRobustCommand{\gtrsim}{%
  \mathrel{\mathpalette\lowersim\originalgtrsim}%
}

\makeatletter
\newcommand{\lowersim}[2]{%
  \sbox\z@{$#1<$}%
  \raisebox{-1pt}{\small $\m@th#1#2$}%
}
\def\({\left (}
\def\){\right )}
\def\[{\left [}
\def\]{\right ]}

\def\J{{\cal J}}
\usetikzlibrary{calc}
\numberwithin{equation}{section}

\usepackage{environ}
\NewEnviron{myequation}{%
\begin{eqnarray}
\scalebox{1.05}{$\BODY$}
\end{eqnarray}
}
   
\newcommand{\beq}{\begin{equation}}
\newcommand{\eeq}{\end{equation}}
 
\newcommand{\bea}{\begin{eqnarray}}
\newcommand{\ea}{\end{eqnarray}}
\newcommand{\barr}{\!\begin{array}}
\newcommand{\earr}{\end{array}\!}

\usepackage{mciteplus} 
\def\nspc{\hspace{-.5pt}}

\def\HH{{\bf H\smpc}}
\def\ddelta{{{\mbox{\fontsize{7pt}{7pt}${\nspc\Delta}$}}}}
\def\spc{\hspace{1pt}}

\counterwithout{equation}{section}

\def\smpc{{\hspace{.5pt}}}
\usepackage{upgreek}

\usetikzlibrary{decorations.pathreplacing,decorations.markings,snakes}

\def\be{\begin{equation}}
\def\ee{\end{equation}}

\def\la{\langle}
\def\bea{\begin{eqnarray}}
\def\eea{\end{eqnarray}}
\def\is{\!  & \!  = \!  &  \!}
\def\ra{\rangle}

\def\ea{\eea}

\def\be{\bea}
\def\ee{\eea}

\def\tr{{\rm tr}}
\renewcommand\large{\fontsize{13}{13.5}\selectfont}
\def\Aminus{{A_-}} 
\def\Aplus{{A_+}} 

\def\spc{\hspace{1pt}}
\def\is{\! &  \! = \! & \!}

\def\ttau{\mbox{\scriptsize${T}\nspc$}}

\usepackage{titling} 
\usepackage[affil-it]{authblk} 
\begin{document}

\addtolength{\topmargin}{3cm}

\title{\bf Double-scaled SYK, Chords and de Sitter Gravity}

\author{Herman Verlinde}

\affil{ Department of Physics, Princeton University, Princeton, NJ 08544}

\affil{School of Natural Sciences, Institute for Advanced Study, Princeton, NJ 08540}
\date{} 

    \maketitle
\bigskip

\bigskip

    \begin{abstract}
   {We study the partition function of 3D de Sitter gravity defined as the trace over the Hilbert space obtained by quantizing the phase space of non-rotating Schwarzschild-de Sitter spacetime. Motivated by the correspondence with double scaled SYK, we identify the Hamiltonian with the gravitational Wilson-line that measures the conical deficit angle. We express the Hamiltonian in terms of canonical variables and find that it leads to the exact same chord rules and energy spectrum as the double scaled SYK model. We use the obtained match to compute the partition function and scalar two-point function in 3D de Sitter gravity. }
    \end{abstract}

\addtolength{\topmargin}{-3cm}
\pagebreak

\tableofcontents
\addtolength{\baselineskip}{0.5mm}
\addtolength{\parskip}{1mm}
\addtolength{\abovedisplayskip}{.65mm}
\addtolength{\belowdisplayskip}{.65mm}
\pagebreak

\def\qbigrt{\bigr)}
\def\darkblue{blue!90!black}
\def\darkredd{red!90!black}
\def\darkred{black}
\def\darkgreen{green!50!black}
\def\lcyan{cyan!50!white}

\def\lgray{gray!50!white}
\section{Introduction}

\vspace{-1.5mm} 

\def\iseqto{ \!&\!\equiv\!&\! }
\def\mfaa{\alpha}
\def\mfa{\smpc{\mathfrak{a}}}
\def\mfns{\mbox{\footnotesize $\mathfrak{n}$}}
 \usetikzlibrary{patterns}
 
\def\mfq{{\sf q}}

Recent studies  indicate that the high temperature limit of the double scaled SYK model \cite{kitaevTalks,Maldacena:2016hyu,Cotler:2016fpe,Berkooz:2018jqr,Berkooz:2018qkz}  can provide a quantum description of low-dimensional de Sitter space \cite{HVtalks}\cite{Susskind:2021esx,Susskind:2022bia, Susskind:2022dfz, Lin:2022nss}\cite{ustwo}\cite{Rahman:2022jsf,Rahman:2023pgt,Rahman:2024vyg}. In this paper we will put the latter correspondence on firmer footing by establishing a direct link between the combinatorial chord rules used in deriving the spectrum and correlation functions in DSSYK and the quantum properties of 3D de Sitter gravity.

The SYK model is defined by the Hamiltonian for $N$ Majorana fermions with commutators $\{\psi_i,\psi_j\} = 2\delta_{ij}$ with a p-body interaction with gaussian random couplings 
\bea
\label{syk}
H \spc = \spc {i}^{p/2}\! \sum_{i_1\ldots i_p}\! J_{i_1\ldots i_p} \psi_{i_1} \ldots \psi_{i_p},  \qquad \  \ \langle  (J_{i_1 \ldots i_p})^2\rangle   =  \frac{\J^2 }{\lambda {N \choose p}}
\eea
We will consider the model in the double scaling limit $N,p \to \infty$ with $\lambda = 2p^2/N$ fixed. As was shown in \cite{Berkooz:2018qkz, Berkooz:2018jqr},  in this limit the computation of moments of the Hamiltonian $H$ reduces to the counting problem of chord diagrams weighted by factor $\mfq^{\# \rm intersections}$ with $\mfq\spc = \spc e^{-\lambda}$. Each chord represents a Wick contraction from the gaussian disorder average of the random couplings. 

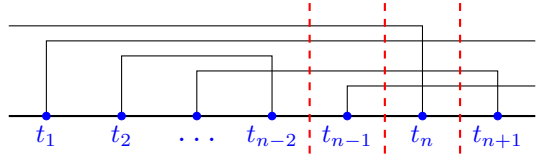
\begin{figure}[hbtp]
 \centering
\begin{tikzpicture}[yscale=1,xscale=1]
\draw[thin] (-2.5,1.2) -- (3,1.2) -- (3,0) ;
\draw[thin] (-2,0) -- (-2,1) -- (4.5,1)  ;
\draw[thin] (-1,0) -- (-1,.8) -- (1,.8) -- (1,0);
\draw[thin] (0,0) -- (0,.6) -- (4,.6) -- (4,0);
\draw[thin] (2,0) -- (2,.4) -- (4.5,.4) ;
\draw[thick] (-2.5,0) -- (4.5,0);
\draw[dashed, red, thick] (1.5,-0.5) -- (1.5,1.6);
\draw[dashed, red, thick] (2.5,-0.5) -- (2.5,1.6);
\draw[dashed, red, thick] (3.5,-0.5) -- (3.5,1.6);
\draw[thin,fill,blue] (-2,0) circle (.5mm) node [below]{\small $t_1$};
\draw[thin,fill,blue] (-1,0) circle (.5mm) node [below]{\small $t_2$};
\draw[thin,fill,blue] (0,0) circle (.5mm) node [below]{};
\draw[thin,fill,blue] (0,-.15) circle (0mm) node [below]{\large$\ldots$};
\draw[thin,fill,blue] (1,0) circle (.5mm) node [below]{\small $t_{n-2}$};
\draw[thin,fill,blue] (2,0) circle (.5mm) node [below]{\small $t_{n-1}$};
\draw[thin,fill,blue] (3,0) circle (.5mm) node [below]{\small $t_{n}$};
\draw[thin,fill,blue] (4,0) circle (.5mm) node [below]{\small $t_{n+1}$};
\end{tikzpicture}
\vspace{-2mm}
\caption{Chord diagram produced by contractions between $H$ insertions at successive time steps.}
\end{figure}

By slicing open a chord diagram, one obtains a combinatorial basis of Hilbert states labeled by the number of chords \cite{Berkooz:2018jqr,Lin:2023trc}. For the partition function, there is only one type of chords. Following the chord rules, the combinatorial Hamiltonian $\HH =\lambda^{1/2} H$ acts on the eigen states $|n\ra$ of the chord number as follows \cite{Berkooz:2018jqr}
 \bea
 \label{hamrule}
 \HH|\smpc n\smpc \ra \! \is \! | n\nspc+\nspc1\ra + [n]_\mfq \, |\smpc n\nspc-\nspc 1\ra, \quad\qquad  [n]_\mfq \equiv\frac{1-\mfq^{n}\!}{1-\mfq}.
\eea
This rule is derived as follows \cite{Berkooz:2018jqr}. Consider a sequence of time steps defined by successive insertions of the Hamiltonian $\HH$. Suppose there are $n$ open chords at a given stage. Acting with $\HH$ at the next stage can either add  a new chord or remove an existing chord via a Wick contraction. The new chord is prescribed to have no intersections with the other chords. A chord that is closed via the Wick contraction with the Hamiltonian insertion, on the other hand, can only do so by first intersecting all the chords between itself and the new Hamiltonian insertion. Summing over all the possible contractions gives $n$ contributions with increasing powers of $\mfq$ from $\mfq^0$ to $\mfq^{n-1}$. Using the geometric series formula gives the above formula.

The recursive formula \eqref{hamrule}, when combined with its generalization to include matter chords, propels the exact solution of DSSYK correlation functions \cite{Berkooz:2018jqr}. Our goal is to derive the same formula from the quantization of 3D Schwarzschild-de Sitter spacetime. Several pieces of evidence indicate that such a derivation should be possible. First, the rule \eqref{hamrule} can be expressed in terms of $\mfq$-deformed oscillators and leads to a chord Hilbert space that admits the action of a $U_{\mfq}(\mathfrak{sl}_2)$ quantum group symmetry. Pure 3D de Sitter gravity is an exactly soluble theory and admits a first order formulation terms of $SL(2,\mathbb{C})$ Chern-Simons theory \cite{Witten:1989ip} with the same quantum symmetry. Notably, both quantum symmetries have the same real deformation parameter $\mfq$ between $0$ and $1$. 

 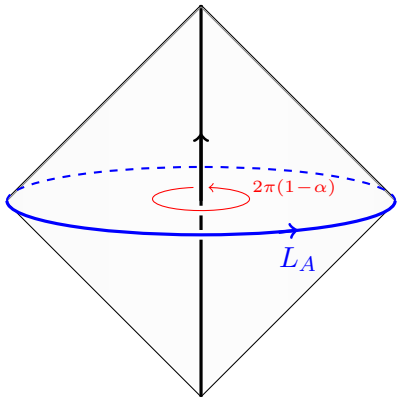
\begin{figure}[t]
 \centering
 \begin{tikzpicture} 
 \begin{scope}[yscale=1.29,xscale=1.29]
 \tikzset{
    partial ellipse/.style args={#1:#2:#3}{
        insert path={+ (#1:#3) arc (#1:#2:#3)}
        }
 }
 \fill[
  left color=darkgray,
  right color=darkgray,
  middle color=white,
  shading=axis,
  opacity=.01
  ] 
  (2,0) -- (0,2) -- (-2,0) arc (180:360:2cm and 0.35cm);
 \fill[
  left color=darkgray,
  right color=darkgray,
  middle color=white,
  shading=axis,
  opacity=.01
  ] 
  (2,0) -- (0,-2) -- (-2,0) arc (180:360:2cm and 0.35cm);
 \draw[very thick,black] (0,-2.02) -- (0,-.4); 
 \draw[very thick,black] (0,-.3) -- (0,-.15);
 \draw[very thick,black] (0,-.05) -- (0,2.02);
 \draw[->,red] (0,0.02)  [partial ellipse=99:441:.5cm and .12cm] node[right] {\quad\,\tiny $2\pi(1\!-\! \alpha)$} ;
\draw[thin,black] (0,2.02) -- (2.0,0.02);
\draw[thin,black] (0,2.02) -- (-2.0,0.02);
\draw[thin,black] (0,-2.02) -- (2.0,-0.02);
\draw[thin,black] (0,-2.02) -- (-2.0,-0.02);
 \draw[gray] (0,0)    (2,0) -- (0,2) -- (-2,0) arc (180:360:2cm and 0.35cm);
  \draw[very thick,->,black] (0,0) -- (0,.7);
\draw[thick,blue,dashed]  (2,0)    arc (0:180:2cm and 0.35cm);
\draw[very thick,blue]  (-2,0)    arc (180:360:2cm and 0.35cm);
 \draw[very thick,->,blue] (0.8,-.325) -- (1,-.3);  
 \draw[blue] (1,-.6)  node {$L_A$};
 \end{scope}
 \end{tikzpicture}
 \vspace{-2mm}
 \caption{The gravitational Wilson line $L_A$ wraps the cosmological horizon and measures the deficit angle of the Schwarzschild-de Sitter spacetime. We identify $L_A$ with the de Sitter Hamiltonian.}
 \vspace{-1mm}
  \end{figure}

A second more physical hint is that the DSSYK energy spectrum derived from \eqref{hamrule} is bounded and naturally parametrized by a spectral angle $\theta$ via
\bea
\HH |\spc \theta \spc \ra \!\is\! \frac{\cos\theta}{\sqrt{1\nspc-\nspc \mfq}} \spc | \spc \theta \spc\ra
\eea
3D Schwarzschild-de Sitter space also has a bounded spectrum characterized by a deficit angle $2\pi\alpha$. In the quantum gravity theory, this angle is measured by a gravitational Wilson line operator $L_A$, as indicated in Figure 2. This line operator $L_A$ will be our main object of study. Its eigen states 
\bea
L_A |\spc \alpha \spc \ra \!\is\! 2\sin(\pi \alpha)\spc | \spc \alpha \spc\ra
\eea
represent Schwarzschild-de Sitter (SdS) spacetimes with a given total mass. Given this, and the similarity between the DSSYK and SdS spectra, it is natural to look for a holographic relationship between the two systems by identifying the Hamiltonian of the gravity theory with the line operator
\bea
\label{hident}
\HH \! \is \! \frac{L_A}{\sqrt{1-\mfq}}
\eea
and by postulating that the spectral angle $\theta$ and deficit angle  $2\pi\alpha$ are related via $2\pi \alpha = \pi-2\theta$.

In the following we will take this perspective to its logical conclusion and explicitly compute the eigen spectrum of the gravitational Wilson line $L_A$. We find that the recursive formula \eqref{hamrule} arises as a direct consequence of the familiar skein relations and other known results in $SL(2,\mathbb{C})$ Chern-Simons theory and pure 3D (A)dS gravity. The identification between the DSSYK and de Sitter parameters is uniquely fixed by matching the $\mfq$-oscillator algebra that appears on both sides.

 \pagebreak
 
 \section{Quantizing 3D Schwarzschild-de Sitter}
 
\vspace{-1.5mm}

In this section, we will develop a Hamiltonian quantum treatment of 3D Schwarzschild-de Sitter spacetime. To set notation, we start with a recap of some properties of the classical spacetime.

\subsection{Symmetries} 
\vspace{-1.5mm}

Global 3D de Sitter spacetime is a maximally symmetric lorentzian submanifold of $\mathbb{R}^{3,1}$ specified a hyperboloid $
-X_0^2 + X_1^2 +   X_2^2 + X^2_3 =1$. The group of isometries acting on 3D de Sitter spacetime is isomorphic to $SL(2,\mathbb{C})$. To write the isometry transformations, it is convenient to combine the embedding coordinates into a $2\times 2$ matrix  
\bea
{\bf X}=  \left(\!\!\begin{array}{cc} \!\! X_0\nspc+\nspc X_3\!\!  &\! X_1\nspc+\nspc iX_2 \\[.5mm] X_1\nspc -\nspc iX_2 &\! X_0\nspc -\nspc X_3\end{array}\!\! \right), 
\ \ & & \ \ \det {\bf X} \spc = \spc -1.
\eea
The isometry group acts on the matrix ${\bf X}$ via conjugation
\bea 
 {\bf X} &\to & g_-  \spc {\bf X}\, g_+,\quad \qquad g_+ = g_-^\dag \in SL(2,\mathbb{C}).
 \eea 
 These transformations preserve the embedding equation and reality condition on $X_a$. The global isometry group is a gauge symmetry of the quantum gravity theory. It will play an important role in what follows.

In static coordinates  $(X_0,X_3,X_1,X_2) = \bigl(\sqrt{1\nspc-\nspc r^2}\spc \sinh t,  \sqrt{1\nspc-\nspc r^2}\spc \cosh t,  r \cos \varphi, r \sin\varphi\bigr),$
the de Sitter metric takes the familiar form 
\bea
\label{statmetric}
ds^2  \! \is\! -(1-  r^2) dt^2 + \frac{dr^2}{1-  r^2} + r^2 d\varphi^2.
\eea
Schwarzschild-de Sitter spacetime is a generalization of global de Sitter with a localized matter source.
 Without loss of generality, we assume the matter source is placed at the origin of the static patch.
The 3D Schwarzschild-de Sitter (SdS) spacetime takes the same form as the vacuum solution \eqref{statmetric}, but with a modified periodic identification of the angular coordinate~$\varphi$
\bea
\label{deficitangle}
\varphi \equiv \varphi + 2\pi (1\nspc-\nspc\mfaa), \ & &\  1\nspc -\nspc\mfaa = \sqrt{1\nspc  -\nspc 8G_N M}\; < \; 1.
\eea
The localized matter source thus creates a conical singularity with deficit angle $2\pi \mfaa$.

A spatial slice of the static patch is a disk bounded by the circle $r=1$,  the cosmological horizon of the observer at $r=0$. The $X_0=0$ spatial slice of global de Sitter, on the other hand, is the two-sphere $X_1^2 + X_2^2 + X^2_3 =1$. 
The static coordinates only span the region $X_1\geq 0$ and thus cover only on the northern hemisphere. The southern hemisphere $X_1\leq 0$ describes a separate static patch given by the causal wedge of an observer at the south pode. The north and south patch are causally disconnected from the observer sitting at the opposite pode.

A static coordinate system that covers that both the north and south patch is 
\bea
(X_0,X_3,X_1,X_2) \!\is \!  (\cos\rho \sinh t, \cos \rho \cosh t,  \sin\rho\cos \varphi, \sin \rho \sin\varphi)
\eea
 with $t\in\mathbb{R}$,  $0\leq \varphi \leq 2\pi(1-\alpha)$  and $\ 0\leq \rho\leq \pi$. In these coordinates the metric becomes
\bea
\label{hopfm}
ds^2 \! \is \! - \cos^2\nspc\nspc \rho\, d t^2 + \sin^2\nspc\nspc \rho \, d\varphi^2 + d\rho^2
\eea
The $t=0$ spatial slice covers a full two-sphere with a north and south pode located at $\rho=0$ and $\rho=\pi$. The hemispheres $\rho\leq \pi/2$ and $\rho\geq \pi/2$ are separated by the observer horizon located at $\rho = \pi/2$ and  are interchanged by the parity transformation that maps $\rho$ to $\pi-\rho$. Note that both podes see the same equator and deficit angle $\alpha$. We will introduce a ${\cal{CPT}}$-symmetry requirement that forces the mass on the north and south patch to be identical.

Quantum field theory on 3D de Sitter spacetime is invariant under ${\cal{CPT}}$-transformations.
In quantum gravity,  this symmetry must be gauged \cite{Harlow:2023hjb}. Motivated by the correspondence with DSSYK, we will now choose a special implementation of ${\cal{CPT}}$-invariance that will break the global $SL(2,\mathbb{C})$ isometry group down to $SU(1,1)$, as follows.
Let $J$ denote the $2\times 2$ matrix
\bea
J \! \is \! \left(\begin{array}{cc} \! 1\! &\! 0 \\[-.0mm] \! 0\! &\! -1 \end{array}\!\! \right) 
\eea
We then define the parity operation ${\cal P}$ and time reflection symmetry ${\cal T}$ via
\bea
{\cal P}: \qquad\,  \mathbf{X} \, \,\to\,- J\spc \mathbf{X}J  \quad & & \ \ \ \; (\rho,\varphi, t) \,\to\, (\pi\nspc - \nspc \rho, \varphi, t\spc)\\[2mm]
{\cal T}: \qquad\;   \mathbf{X} \,\to\, -\mathbf{X}^{-1} \quad\;  & & \ \quad (\rho,\varphi, t) \,\to\,  (\smpc\rho\smpc,\spc \varphi\smpc,-t\spc)
\eea
The parity operation ${\cal P}$ interchanges the north and the south patch. The combined transformation
\bea
\label{pete}
{\cal P}{\cal T}: \qquad\,  \mathbf{X} \, \to\, J\spc \mathbf{X}^{-1}\nspc J  \quad\  & & \quad\  (\rho,\varphi, t) \,\to\,  (\pi\nspc - \nspc \rho, \varphi, -t\spc)
\eea
interchanges the two static patches, while reversing the direction of time. In terms of the embedding coordinates, the ${\cal P}{\cal T}$ operation flips the sign of the $X_3$-coordinate. Gauging ${\cal{CPT}}$ symmetry requires that all physical states and operators must be invariant under the combined operation of flipping the $X_3$ coordinate and applying the charge conjugation map ${\cal C}$ to all matter fields.

The isometries that commute with the ${\cal P}{\cal T}$ are given the $SU(1,1)$ subgroup of $SL(2,\mathbb{C})$ transformations specified by the restriction\footnote{One directly verifies that  $g^\dag (J\smpc \mathbf{X}^{-1}) J g  = J  (g^\dag\mathbf{X}g)^{-1} J$ and $\tr(hgh^\dag J)  = \tr(gJ)$ for all $g,h\in SU(1,1)$ .}
\bea
\label{suoneone}
g^\dag J g \!\is\! J. 
\eea 
For later reference, we note that $SU(1,1)$ admits two invariant traces, the usual one and${}^{1}$
\bea
\label{tracej}
&  \tr_J(g) \spc \equiv -i\; \tr(gJ) &
\eea

\pagebreak

\subsection{Phase space}
\vspace{-1mm}

In the quantum gravity theory, the deficit angle $\alpha$ becomes an operator. It will be useful to formulate the Schwarzschild-de Sitter spacetime as a quotient of global de Sitter. The matter source traces out a worldline ${\cal C} = \{ X_2=X_3=0\}$. De Sitter spacetime with ${\cal C}$ removed is non-simply connected. Let $A$ denote the non-contractible homology cycle that surrounds ${\cal C}$. The SdS spacetime is then specified by the following holonomy condition on the embedding coordinates
\bea
\label{geea}
\qquad\qquad {\bf X} \to g_A \spc {\bf X} \spc g_A^{-1},\ \  & &\ \   g_A= \left(\begin{array}{cc} \!\! {e^{{i\pi \alpha}}}\!\!  &\! 0\! \\[-.0mm] 0 &\! {e^{-{i\pi\alpha}}}\!\end{array}\!\!\right) \qquad\qquad \Bigl(\spc\raisebox{-6pt}{$\stackrel{\mbox{\footnotesize A-cycle}}{\mbox{\footnotesize holonomy$\strut$}}$}\Bigr)\qquad
\eea
The coordinate matrix ${\bf X}$ becomes multivalued and undergoes an isometry transformation $g_A$ when going around the trajectory ${\cal C}$. 
Since physical observables are required to be invariant under the global isometry group, the holonomy $g_A$ itself is not measurable, but its trace 
\bea
L_A \!\! & \! \equiv \! & \!\! \tr_J(\spc g_A) \, = \, 2\sin(\pi \alpha)
\eea
is a measurable quantity. Here we used the special $SU(1,1)$ invariant trace defined in \eqref{tracej}. The variable $L_A$ uniquely specifies the Schwarzschild-de Sitter geometry up to global isometries. 

The phase space associated with the Schwarzschild-de Sitter spacetime is two-dimensional. The second phase space coordinate is found by considering the embedding of the static patch within the global de Sitter spacetime.  Let $t_N$ denote the time coordinate on the northern patch and $t_S$ the time coordinate on the southern patch. Since the north and south patches are causally disconnected, it is natural to view the two time coordinates defined on each patch as independent. In general, the two time coordinates are related via a non-trivial time-shift
\bea
t_N \! \is \! t_S + z
\eea
We can view this time shift as a transition function that relates the two coordinate charts on $\rho\leq\pi/2$ and $\rho\geq\pi/2$. In terms of the matrix coordinate ${\bf X}$, the transition function is represented via the isometry transformation
\bea
\label{ztrans}
\qquad\quad {\bf X} \  \to \ h_Z \spc {\bf X} \spc h_Z, \ & & \ \ h_Z\spc  = \spc \left(\begin{array}{cc} \!\! {e^{{z/2}}}\!\!  &\! 0 \\[-.0mm] 0 &\! {e^{-{z/2}}}\end{array}\!\! \right) \qquad\quad \Bigl(\,\raisebox{-6pt}{$\stackrel{\mbox{\small North-South}}{\mbox{\small holonomy$\large\strut$}}$}\Bigr)
\eea
The timeshift parameter $z$ is the other phase space variable associated with the Schwarzschild-de Sitter space time. 

Our next goal is to determine the Poisson bracket between the classical phase space variables $\alpha$ and $z$ and find their quantum realization. Fortunately, modulo some straightforward modifications,  most of this work has already been done for us \cite{Chekhov:1999tn,Teschner:2003em,Terashima:2011qi}. Before turning to this task, we first give a second description of the phase space of the Schwarzschild-de Sitter spacetime that is better adapted for constructing the quantum theory.



\def\scL{{\sc L}}
\def\sfr{{\mathfrak{s}}}

To set up the quantum theory of the SdS spacetime, it is convenient to transition to a first order formulation in terms of a triad $e^a$ and $SO(2,1)$ spin connection $\omega^a$  \cite{Witten:1989ip}.  The linear combinations $A^a_{\pm} = \omega^a \pm i e^a$
constitute two $SL(2,\mathbb{C})$ connections related via complex conjugation $A^a_+ = (A^a_-)^*$. 
In terms of these variables, the 3D Einstein action takes the form of an $SL2,\mathbb{C})$ Chern-Simons actions with imaginary coupling constant $k = i\kappa$ 
\bea
\label{csaction}
S\!  \is \! \frac{i\kappa}{4\pi} \! \int\! {\rm Tr} \Bigl(\Aplus d\Aplus + \frac 2 3 \Aplus\nspc\wedge\nspc\Aplus\nspc\wedge\nspc\Aplus 
\Bigr) - \frac{i\kappa}{4\pi} \! \int\! {\rm Tr} \Bigl( \Aminus d\Aminus + \frac 2 3 \Aminus\nspc\wedge\nspc\Aminus\nspc\wedge\nspc\Aminus 
\Bigr)^{\strut} \nonumber \\[-1mm]\\[-1mm]\nonumber
& & \qquad\ \  \ \kappa \, = \, \frac{1}{2G_N}, \qquad \ \Aplus{\!\!\!}^a\, \, = \, \omega^a + i e^a, \qquad \Aminus{\!\!\!}^a\, \, = \, \omega^a - i e^a
\eea 
The action \eqref{csaction} defines a topological QFT and exhibits diffeomorphism invariance as well as $SL(2,\mathbb{C})$ gauge symmetry. This first order formulation of 3D gravity has the advantage that it gives a simple characterization of the phase space of classical solutions in the presence of localized matter sources and leads to a convenient expression for the symplectic form on this phase space that facilitates the derivation of the canonical commutation relations in the quantum theory.

In the absence of any matter source, the  Einstein equation and torsion constraint derived from \eqref{csaction} take the form of flatness constraints
\bea
\label{flat}
F(\Aplus) \! \is \! F(\Aminus) \spc = \spc 0
\eea
that imply the constant curvature equations for the metric, along with the torsion constraints that allows one to solve for the spin connection $\omega_a$ in terms of the triad $e_a$. In this way, the phase space of classical solutions of de Sitter gravity mapped to the space of solutions to the flatness conditions \eqref{flat} modulo $SL(2,\mathbb{C})$ gauge transformations \cite{Witten:1989ip}.

Localized matter sources like massive point particles are described by $SL(2,\mathbb{C})$ Wilson lines  along the wordline evaluated in an appropriate representation specified by the mass and spin or angular momentum. For light particles of mass squared $m^2<1$, the Wilson line is evaluated in a complementary series representation with real $SL(2,\mathbb{C})$ spin  $0\leq \Delta\leq 1$ related to $m$ via
\bea
m^2 \! \is\! 4\Delta(1-\Delta) 
\eea
For heavy particles with $m^2>1$, the Wilson line is taken in a continuous series representation labeled by a complex spin $\Delta = \frac 1 2 + is$. We will only consider non-rotating matter sources.

In the presence matter, the flatness conditions \eqref{flat} are imposed only outside of the localized matter sources. The trace of the holonomy of the gauge fields $A_\pm$ around the matter sources define gauge invariant observables that specify the phase space parameters of the gravity theory such as the angle deficit, timeshifts, and/or the geodesic distance between of the matter source. In the Chern-Simons formalism, the holonomy and the deficit angle $2\pi \alpha$ reflect the fact that the matter Wilson lines at the north and south pode create curvature singularity proportional to the mass $m$. For small mass $m$, the deficit angle is given by $2\pi \alpha = \frac{2\pi}{\kappa} m = 4G_Nm$.
 For non-rotating matter sources, the trace of the holonomy of the $A_+$ and $A_-$ are equal.

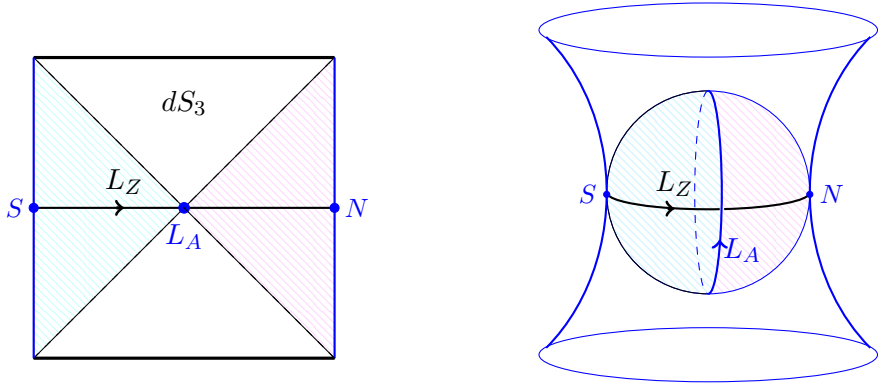
\begin{figure}[t]
\begin{center}
${}$~~~~~~~~~~~\begin{tikzpicture}[scale=1]
\draw (4,1.4) node {$dS_3$};
\draw[very thick] (2,-2) -- (6,-2);
\draw[very thick] (2,2) -- (6,2);
\draw[dashed] (2,-2) -- (6,2);
\draw[fill, pattern={north west lines}, pattern color={cyan!20!white}]  (2,2) -- (4,0) -- (2,-2) -- (2,2);
\draw[fill, pattern={north west lines}, pattern color={magenta!15!white}]  (6,2) -- (4,0) -- (6,-2) -- (6,2);
\draw[thick,blue] (6,-2) -- (6,2);
\draw[thick,blue] (2,-2) -- (2,2);
\draw[thick] (2,0) -- (6,0);
 \draw[very thick,->] (3.1,0) -- (3.2,0); 
\draw[thin,fill,blue] (2,0) circle (.6mm) node [left]{\small $S$};
\draw[thin,fill,blue] (6,0) circle (.6mm) node [right]{\small $N$};
\draw[thin,fill,blue] (4,0) circle (.7mm) node [below]{$\strut L_A$};
\draw[thin,fill] (3.2,0.05) circle (0mm) node [above]{$L_Z$};
\draw[dashed] (6,-2) -- (2,2);
\end{tikzpicture}~~~~~~\raisebox{-3mm}{\begin{tikzpicture}[rotate=90]
\begin{scope}[yscale=.9,xscale=.9]
 \tikzset{
    partial ellipse/.style args={#1:#2:#3}{
        insert path={+ (#1:#3) arc (#1:#2:#3)}
    }
}       \draw[blue,pattern={north west lines}, pattern color={magenta!15!white}] [thin] (-2,0) circle (1.5cm);
        \draw[pattern={north west lines}, pattern color={cyan!20!white}]  (-2,0) [partial ellipse=0:180:1.5cm and 1.5cm];
        \draw[thick,blue,pattern={north west lines}, pattern color={cyan!20!white}] (-2,0) [partial ellipse=180:360:1.5cm and .2cm];
        \draw[blue,thick] (-2,4)  [partial ellipse=220:320:3cm and 2.5cm];
        \draw[blue,thick] (-2,-4)  [partial ellipse=40:140:3cm and 2.5cm];
        \draw[blue] (.4,0)  [partial ellipse=0:360:.4cm and 2.5cm];
        \draw[blue] (-4.4,0)  [partial ellipse=0:360:.4cm and 2.5cm];
        \draw[blue, dashed] (-2, 0) ellipse (1.5cm and .2cm);
        \node[blue] at (-2.85, -.5) {$L_A$};   
        \draw[very thick,->] (-2.24,0.6) -- (-2.24,0.5); 
         \draw[very thick,->,blue] (-2.74,-0.17) -- (-2.7,-0.17); 
        \draw (-2,0)[thick] [partial ellipse=90:186.5:.25cm and 1.5cm];
        \draw (-2,0)[thick] [partial ellipse=189:270:.25cm and 1.5cm]; 
        \draw[thin,fill,blue] (-2.03,1.5) circle (.45mm) node [left]{\small $S$};
        \draw[thin,fill,blue] (-2.03,-1.5) circle (.45mm) node [right]{\small $N$};      
         \draw[thin,fill] (-2.2,.5) circle (0mm) node [above]{$L_Z$};
        \end{scope}
\end{tikzpicture}}
\vspace{-2mm}
\end{center}
\caption{3D de Sitter space with the holonomy variables $L_A$ and $L_Z$ indicated. The left figure shows the Penrose diagram with the two static causal wedges separated by the cosmological horizon, the right figure depicts the spatial constant time slice, which takes the form of a three sphere divided into two hemispheres. In the first order formulation of 3D gravity, the holomies $L_A$ and $L_Z$ are Wilson lines of a flat $SL(2,\mathbb{C})$ connection along two dual cycles with a single intersection point. }
\end{figure}

As an explicit and relevant example, the triad and spin connection that specify the Schwarzschild-de Sitter metric \eqref{hopfm} in Hopf coordinates are given by \cite{Castro:2011xb}
\bea
&\omega\!&\!\! = \spc \omega^a\spc \ttau_a \spc = \spc  i \sin\nspc \rho\,  \ttau_3 \spc dt + \cos\nspc \rho \, \ttau_2\spc d\varphi \nonumber \\[-2.5mm]\\[-2.5mm]\nonumber
e \!\is \!   e^a\spc \ttau_a \spc = \spc \ttau_1 \spc d\rho + \cos\nspc \rho\, \ttau_2\spc dt + i  \sin\nspc \rho \, \ttau_3\spc  d\varphi 
\eea
These one forms assemble into a complex $SL(2,\mathbb{C})$ connection ${\cal A} = (A_+,A_-)$  with non-trivial holonomy around the A-cycle (which choose to locate at the equator $\rho=\frac{\pi}2$)
\bea
\label{apm}
A_\pm  = \spc A_\pm^a\, \ttau_a\!\!  \is\!
\pm  i\ttau_1 \spc d\rho \pm i  \ttau_3 \spc (d t \pm i d\varphi)  \\[2mm] 
g_A^\pm \spc \equiv\spc  {\rm P}\exp\! \oint_A\! A_\pm \!\!\!\!\!\!\!\!\!\! & &\!\!\!= \cos(2\pi \alpha) +i  \sin(2\pi \alpha)\spc   \ttau_3\spc  .
\eea
For non-rotating Schwarzschild-de Sitter solutions, the gauge invariant holonomy variables associated with the A-cycle are both equal
\bea
L_A^+ \spc =\spc L^-_A  
 \! \is \! 2 \sin(\pi \alpha), \qquad \qquad L^\pm_A \spc \equiv \spc \tr_J(g_A^\pm) 
\eea
Here the holonomy is evaluated in the spin 1/2 representation and via the special $SU(1,1)$ invariant trace \eqref{tracej}.
From now on, we will restrict the gravity theory to its s-wave sector by imposing the condition $L_A^+ \spc =\spc L^-_A$ as a gauge constraint that eliminates all states with non-zero angular momentum.

We can also compute the path-ordered exponential of the $SL(2,\mathbb{C})$ connection evaluated along a straight path between the north and south pode. Using that the north and south patch are glued together via the transition function \eqref{ztrans}, we find that
 \bea
g_Z^\pm \spc \equiv\, \spc {\rm P}\exp\int_{S}^{N} \!\! A_\pm \! \is \!   \left(\begin{array}{cc} \!\! { 0} &\! \!{e^{{z/2}}}\\[-.0mm] \!\! - {e^{-{z/2}}}\!\! & 0 \end{array}\!\! \right)
\eea
The conjugacy class of the north-south holonomy $g^\pm_Z$ is gauge invariant under global $SL(2,\mathbb{C})$ transformations that act identically at the two podes.  As the second gauge invariant phase space variable of the Schwarzschild-de Sitter spacetime we choose the top component of $g_Z$\footnote{This choice amounts to a Hamiltonian reduction procedure.}
\bea
L_Z^+ \spc =\spc L^-_Z  
 \! \is \!  e^{z/2} 
\eea
The variable $z$ can be thought of as the time difference between the two antipodal clocks. In the next subsection, we will construct the quantum analogs of the phase space variables $L_A$ and $L_Z$ and compute their commutator algebra and eigen spectrum.
 
Before we turn to the quantum theory, a brief comment is in order.  Above we specified the SdS geometry by means of the holonomy $g_A$ labeled by the deficit angle $2\pi \alpha$. When viewed as a $SL(2,\mathbb{C})$  group element, the angle $2\pi \alpha$ represents a periodic variable defined modulo $2\pi$. A rotation over a multiple of $2\pi$ is trivial. On the geometry side, however, there is a clear physical difference between a Schwarzschild-de Sitter spacetime with zero deficit angle and one with deficit angle equal to $2\pi$. In the following, we will adopt the notion of large gauge transformations suggested by the first order formalism and treat the deficit angle $2\pi \alpha$ as a periodic variable. On the metric side, the state with given $\alpha$ should then be viewed as a linear superposition of geometries with angle deficit (or excess) equal to $2\pi \alpha$ plus any integer multiple of~$2\pi$. 

\subsection{Quantization}
\vspace{-1.5mm}

Our next task is to turn the phase space variables $L_A$ and $L_Z$ into hermitian quantum observables acting on a suitable Hilbert space. For this, we first need to specify the Poisson bracket between the holonomy variables. The first order action \eqref{csaction} is particularly useful for this purpose. 

Let $A^\pm = A^\pm_{\alpha a}\, \ttau^a\spc dx^\alpha$ specify gauge connections living on the 2D spatial slice $\Sigma$. The time components $A_t^\pm$ act as lagrange multipliers that impose the flatness constraints $F(A_\pm) = 0$.  Via the standard rules of canonical quantization, we read off that the symplectic form is given by
\bea
\label{omega} 
\Omega \! \is \! \frac{i\kappa}{4\pi} \int_\Sigma \bigl(\smpc\tr( \delta A^+\! \wedge \delta A^+) -
 \tr\bigl( \delta A^-\! \wedge \delta A^-) \spc\bigr)
\eea
Here $\delta A_\pm$ represent one-forms on the space of flat $SL(2,\mathbb{C})$-connections, modulo gauge transformations. 

We can now follow one of two approaches (i) we can first use the above formulas to quantize the space of all 2D gauge connections and then impose the flatness constraints while dividing out gauge transformations, or (ii) we can aim to directly quantize the finite dimensional phase space of flat connections modulo gauge transformations. Here we will take a hybrid approach: we will use the `quantize first' method to compute the Poisson bracket between the gauge invariant holonomy variables $L_A$ and $L_Z$ and then use this result to directly construct a Hilbert space representation of the operator algebra of physical observables in the full quantum theory.

The symplectic form \eqref{omega} leads to the following local commutators between the gauge fields
\bea
\label{acom}
\bigl[A^\pm_{\alpha a}(x),A^\pm_{\beta b}(y)\spc\bigr]\! \is \! \pm \hbar\spc \epsilon_{\alpha\beta}\spc \delta_{ab} \spc \delta(x-y), 
\qquad \quad \hbar \equiv \frac{2\pi i}{\kappa}
\eea
These are the familiar commutators from quantum Chern-Simons theory, except that the level $k = i\kappa$ is now imaginary, so the $\hbar$ on the right-hand side has an extra factor of $i$. 

In the `quantize first' approach, one first introduces a Hilbert space spanned by wave functionals of a maximally commuting subset of gauge field components and then implements the flatness conditions by requiring that physical states are annihilated by the operators $F(A^+)$ and $F(A^-)$, while imposing that physical operators must commute with the constraints. The physical state conditions automatically implement gauge and diffeomorphism invariance, leaving the holonomies around the non-trivial cycles of the spatial slice $\Sigma$ as the only physical, gauge and diffeomorphism invariant observables.\footnote{This is not quite true. For general Wilson line configurations, diffeomorphism invariance is in fact broken by the framing anomaly. This anomaly cancels in suitable spinless combinations of Wilson lines of the $A^+$ and $A^-$ gauge field. We will return to this point later.}

 Defining the Poisson bracket as the classical limit $\{\ , \ \}_{{\nspc}_{\rm PB}}\!\nspc= \raisebox{-1pt}{${}^{\raisebox{-2pt}{\footnotesize $\lim$}}_{{}_{\hbar\to 0}}$}\, \mbox{\large $\frac 1 \hbar$}\, [\ , \ ]$ of the commutator \eqref{acom}, we can now compute the bracket between the holonomy variables  $L_A$ and $L_Z$. This bracket is non-zero because the two lines intersect \cite{goldman1984symplectic}. From \eqref{acom} it is straightforward to derive \footnote{To avoid clutter and unnecessary repetition, we will from now on drop the $\pm$ superscripts.} 
\begin{figure}[t]
\centering
 \begin{tikzpicture}[rotate=0]
 \begin{scope}[yscale=.84,xscale=.84]
 \tikzset{
    partial ellipse/.style args={#1:#2:#3}{
        insert path={+ (#1:#3) arc (#1:#2:#3)}
    }
}      
           \node[gray!20!black] at (-.37, .25) {\footnotesize $S$};
           \node[gray!20!black] at (2, .25) {\footnotesize $N$};     
           \node[black] at (.53,-.35) {$L_Z$};
        \node[blue] at (2, 1.35) {$L_A$};
         \draw[thick,fill] (-.35,0) circle (.5mm);    
       \draw[thick,fill] (2,0) circle (.5mm);
        \draw[thick,blue] (2,0) [partial ellipse=-180:180:.98cm and .98cm];
        \draw[thick][black] (-.35, 0) -- (.98, 0);
        \draw[thick][black] (1.06, 0) -- (2, 0);
        \end{scope}\end{tikzpicture}~~~~~~~~~~~~~\begin{tikzpicture}[rotate=0]
\begin{scope}[yscale=-.82,xscale=.82]
 \tikzset{
    partial ellipse/.style args={#1:#2:#3}{
        insert path={+ (#1:#3) arc (#1:#2:#3)}
    }
}       
           \node[gray!50!black] at (-.3, -.25) {\footnotesize $S$};
           \node[gray!50!black] at (2, -.25) {\footnotesize $N$};
        \node[cyan!50!blue] at (2, -1.3) {$L_Y$};
       \draw[thick,cyan!60!blue] (.63,-.45) [partial ellipse=15:90:4.5mm and 4.5mm]; 
        \draw[thick,cyan!60!blue] (1.5,0.4) [partial ellipse=140:270:4mm and 4mm];
        \draw[thick,cyan!60!blue] (2,0) [partial ellipse=-159.6:0:1cm and .95cm]; 
        \draw[thick,cyan!60!blue] (1.98,0) [partial ellipse=-0:142:1.02cm and  1.025cm];
        \draw[thick,cyan!60!blue] (-.3, 0) -- (.65, 0);     
        \draw[thick,cyan!60!blue](1.47, 0) -- (2, 0);   
        \draw[thick,fill] (-.3,0) circle (.5mm);    
       \draw[thick,fill] (2,0) circle (.5mm);        \end{scope}
\end{tikzpicture}~~~~~~~~~~~~\begin{tikzpicture}[rotate=0]
\begin{scope}[yscale=.82,xscale=.82]
 \tikzset{
    partial ellipse/.style args={#1:#2:#3}{
        insert path={+ (#1:#3) arc (#1:#2:#3)}
    }
}       
          \node[gray!50!black] at (-.3, .25) {\footnotesize $S$};
           \node[gray!50!black] at (2, .25) {\footnotesize $N$};
        \node[green!60!black] at (2,1.39) { $L_{\,\widetilde{\!Y\nspc}}$};
          \draw[thick,green!60!black] (.63,-.45) [partial ellipse=15:90:4.5mm and 4.5mm]; 
        \draw[thick,green!60!black] (1.5,0.4) [partial ellipse=140:270:4mm and 4mm];
        \draw[thick,green!60!black] (2,0) [partial ellipse=-159.6:0:1cm and .95cm]; 
        \draw[thick,green!60!black] (1.98,0) [partial ellipse=-0:142:1.02cm and  1.025cm];
        \draw[thick,green!60!black] (-.3, 0) -- (.65, 0);     
        \draw[thick,green!60!black](1.47, 0) -- (2, 0);   
        \draw[thick,fill] (-.3,0) circle (.5mm);    
       \draw[thick,fill] (2,0) circle (.5mm);
        \end{scope}
\end{tikzpicture}
\caption{The holonomy variables $L_A$ and $L_Z$ that span the phase space of Schwarzschild-de Sitter (left) and the holonomies $L_Y$ and $L_{\tilde Y}$ that arise in the Poisson bracket between $L_A$ and $L_Z$ (right).}
\vspace{-3mm}
\end{figure}
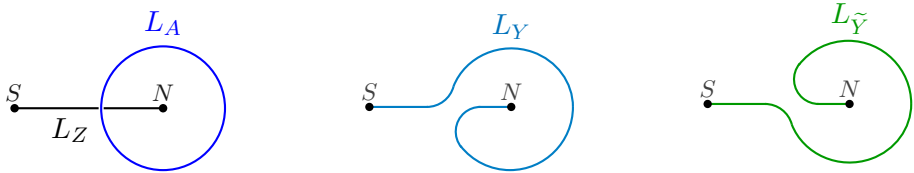
\bea
\label{goldman}
\{L_A,L_Z\}_{\rm PB} \! \is \! L_Y - L_{\tilde Y} 
\eea
where $L_Y$ and $L_{\tilde Y}$ denote the holonomies along the two paths obtained by breaking open the lines $L_A$ and $L_Z$ and reconnecting the two open ends in the two possible ways, as indicated in figure~2. 


Equation \eqref{goldman} is a special case of the general bracket relation between spin 1/2 Wilson lines in $SL(2,\mathbb{C})$ Chern-Simons gauge theory known as the skein relation \cite{Witten:1988hc}\cite{kauffman1988new}\cite{reshetikhin1991invariants}\cite{Witten:1989rw}\cite{Gaiotto:2014lma}\cite{roger2014skein}. 
The classical skein relation can be lifted to an exact operator relation in the quantum theory, by combining the fundamental trace identity $\tr(gh^{-1}) + \tr(gh) = \tr(g)\tr(h)$, which holds for any pair of $SL(2,\mathbb{C})$ matrices, with the exponentiated Poisson bracket skein relation \eqref{goldman}. Zooming in on the intersection point between $L_Z$ and $L_A$, the quantum skein relation takes the simple form\cite{kauffman1988new}\cite{Gaiotto:2014lma} \cite{roger2014skein}\\[-8mm]
\be
\label{skein}
\raisebox{0mm}{ \begin{tikzpicture}[rotate=45]
 \begin{scope}[yscale=.9,xscale=.9]
 \tikzset{
    partial ellipse/.style args={#1:#2:#3}{
        insert path={+ (#1:#3) arc (#1:#2:#3)}
    }
}      
        \draw[thick][blue] (0, 0) -- (2, 0);       
        \draw[thick][black] (1, -1) -- (1, -.09);
        \draw[thick][black] (1, .09) -- (1, 1);
        \end{scope}
\end{tikzpicture}}&&\raisebox{6.25mm}{$= \ \ q^{\frac 1 2}$}~~
 \raisebox{-7.5mm}{\begin{tikzpicture}[rotate=45]
\begin{scope}[yscale=.85,xscale=.85]
 \tikzset{
    partial ellipse/.style args={#1:#2:#3}{
        insert path={+ (#1:#3) arc (#1:#2:#3)}
    }
}       
       \draw[thick,cyan!60!blue] (.5,-2.8) [partial ellipse=0:90:12mm and 12mm]; 
        \draw[thick,cyan!60!blue] (2.8,-.5) [partial ellipse=180:270:12mm and 12mm];
        \end{scope}
\end{tikzpicture}}~~~~\raisebox{6.25mm}{$+ \ \ q^{-\frac 1 2}$\!\!\!\!\!\!\!}\raisebox{-.5mm}{
 \begin{tikzpicture}[rotate=45]
\begin{scope}[yscale=-.8,xscale=.85]
 \tikzset{
    partial ellipse/.style args={#1:#2:#3}{
        insert path={+ (#1:#3) arc (#1:#2:#3)}
    }
}       
    \draw[thick,green!60!black] (-.8,-2.6) [partial ellipse=0:90:12mm and 12mm]; 
        \draw[thick,green!60!black] (1.58,-.1) [partial ellipse=180:270:12mm and 12mm];
        \end{scope}
\end{tikzpicture}}\nonumber\\[-18mm]
\eea

\noindent
where\footnote{Here we absorb the one loop shift $k\to k + 2$ into our definition of the Chern-Simons level. The parameter $\kappa$ labeling $q$ in \eqref{qu} receives a similar one-loop shift $\kappa\to \kappa-2$ relative to the bare Newton constant in the first order action \eqref{csaction}. With the one-loop shift included, the skein relation \eqref{skein} is exact to all orders.} 
\bea
\label{qu}
q \! \is \! 
e^{-\frac{2\pi}{\kappa}} \, =\, e^{-4\pi G_N}_{\strut}
\eea
The skein relation plays an important role in the study of knot invariants \cite{kauffman1988new} and their realization as Wilson line expectation values \cite{Witten:1988hc}. Note that the $q$-parameter in our case is a real number between 0 and 1.\footnote{In the application to knot theory, one often takes $q$ to be a $k$-th root of unity, corresponding to Wilson lines in $SU(2)$ Chern-Simons theory at level $k$. For $SL(2,R)$ CS theory, AdS$_3$ gravity, quantum Teichm\"uler theory, and Virasoro CFT, the skein relation \eqref{skein} also holds for spin 1/2 gravitational Wilson lines that measure geodesic lengths \cite{roger2014skein} and loop operators of the (2,1) degenerate field \cite{Gaiotto:2014lma}. In that case $q$ is a complex phase  $q = e^{\frac{12\pi i}{c}} = e^{8\pi i G_N}$ with $c$ the central charge.
In \cite{ustwotwo} we show that the correlators of double scaled SYK match boundary correlators in Virasoro CFT with complex central charge $c = 13 \pm i\gamma$ and real $q$ parameter. The basic structure that underlies this match is that both theories exhibit the same quantum group and skein relations with identical real values of $q$.} In a general $SL(2,\mathbb{C})$ Chern-Simons theory, $q$ can be any complex number. While special things happen at special values of $q$, many properties of the theory with real $q$ can be derived by analytic continuation from the more well-studied case $|q|=1$. 

Zooming out, the global skein relation for the holonomy variables $L_A$ and $L_Z$ reads as follow
\bea
\label{skeint}
\qquad\qquad\qquad L_Z L_A \is q^{\frac 1 2} \spc L_Y\, +\, q^{-{\frac 1 2}}\spc L_{\,\widetilde{\!Y\nspc}}\qquad\qquad\ \Bigl(\spc\raisebox{-6pt}{$\stackrel{\mbox{\small skein\tiny$\strut$}}{\mbox{\small relation$\strut$}}$}\spc \Bigr)
\eea
or in pictures\\[-6mm]
$$
 \begin{tikzpicture}[rotate=0]
 \begin{scope}[yscale=.84,xscale=.84]
 \tikzset{
    partial ellipse/.style args={#1:#2:#3}{
        insert path={+ (#1:#3) arc (#1:#2:#3)}
    }
}       
           \node[gray!50!black] at (-.45, .25) {\tiny $S$};
           \node[gray!50!black] at (2, .25) {\tiny $N$};     
           \node[black] at (.55,-.25) {\footnotesize $L_Z$};
        \node[blue] at (2, 1.25) {\footnotesize $L_A$};
         \draw[thick,fill] (-.45,0) circle (.5mm);    
       \draw[thick,fill] (2,0) circle (.5mm);
        \draw[thick,blue] (2,0) [partial ellipse=-180:180:.98cm and .98cm];
        \draw[thick][black] (-.45, 0) -- (.98, 0);
        \draw[thick][black] (1.06, 0) -- (2, 0);
        \end{scope}
\end{tikzpicture}~~\raisebox{8mm}{$= \ \ q^{\frac 1 2}$}~~
 \begin{tikzpicture}[rotate=0]
\begin{scope}[yscale=-.82,xscale=.82]
 \tikzset{
    partial ellipse/.style args={#1:#2:#3}{
        insert path={+ (#1:#3) arc (#1:#2:#3)}
    }
}       
           \node[gray!50!black] at (-.3, -.25) {\tiny $S$};
           \node[gray!50!black] at (2, -.25) {\tiny $N$};
        \node[cyan!50!blue] at (2.05, -1.24) {\footnotesize $L_Y$};
       \draw[thick,cyan!60!blue] (.63,-.45) [partial ellipse=15:90:4.5mm and 4.5mm]; 
        \draw[thick,cyan!60!blue] (1.5,0.4) [partial ellipse=140:270:4mm and 4mm];
        \draw[thick,cyan!60!blue] (2,0) [partial ellipse=-159.6:0:1cm and .95cm]; 
        \draw[thick,cyan!60!blue] (1.98,0) [partial ellipse=-0:142:1.02cm and  1.025cm];
        \draw[thick,cyan!60!blue] (-.3, 0) -- (.65, 0);     
        \draw[thick,cyan!60!blue](1.47, 0) -- (2, 0);   
        \draw[thick,fill] (-.3,0) circle (.5mm);    
       \draw[thick,fill] (2,0) circle (.5mm);        \end{scope}
\end{tikzpicture}~~~
\raisebox{8mm}{$+ \ \ q^{-\frac 1 2}$}~\raisebox{-.5mm}{ \begin{tikzpicture}[rotate=0]
\begin{scope}[yscale=.82,xscale=.82]
 \tikzset{
    partial ellipse/.style args={#1:#2:#3}{
        insert path={+ (#1:#3) arc (#1:#2:#3)}
    }
}       
          \node[gray!50!black] at (-.3, .25) {\tiny $S$};
           \node[gray!50!black] at (2, .25) {\tiny $N$};
        \node[green!60!black] at (2.05, 1.33) {\footnotesize $L_{\,\widetilde{\!Y\nspc}}$};
          \draw[thick,green!60!black] (.63,-.45) [partial ellipse=15:90:4.5mm and 4.5mm]; 
        \draw[thick,green!60!black] (1.5,0.4) [partial ellipse=140:270:4mm and 4mm];
        \draw[thick,green!60!black] (2,0) [partial ellipse=-159.6:0:1cm and .95cm]; 
        \draw[thick,green!60!black] (1.98,0) [partial ellipse=-0:142:1.02cm and  1.025cm];
        \draw[thick,green!60!black] (-.3, 0) -- (.65, 0);     
        \draw[thick,green!60!black](1.47, 0) -- (2, 0);   
        \draw[thick,fill] (-.3,0) circle (.5mm);    
       \draw[thick,fill] (2,0) circle (.5mm);
        \end{scope}
\end{tikzpicture}} 
$$ 
We wish to use this skein relation to obtain an explicit expression for the holonomy operator $L_A$ in terms of elementary quantum variables with known, simple commutation relations. It will be helpful to borrow and adapt results obtained in the study of quantum Teichm\"uller space~\cite{Chekhov:1999tn,Teschner:2003em,Terashima:2011qi}. 

The holonomy variables introduced above are a generalization to $SL(2,\mathbb{C})$ of the so-called Penner coordinates that parametrize the moduli space of flat $SL(2,\mathbb{R})$ connections on a 2D surface. The Penner construction follows a similar strategy as  above. First, one defines a triangulation of the 2D surface by introducing a set of marked points connected by edges $e$. The Penner coordinates are the hyperbolic lengths $L_e$ of these edges, defined by taking the inner product $L_{ij} =\la X_i,X_j\ra$ of the embedding coordinate vectors associated with each marked end-point. 

In case the marked points sit at a curvature singularity, as in our situation, the definition of the hyperbolic $L_e$ is a bit subtle and requires a regularization procedure. We will not elaborate this point here, except to remark that it can be dealt with by introducing additional small geodesics that surround the punctures and by taking appropriate cross-ratios of the geodesic lengths. These regularized cross-ratios are called Fock coordinates. 

While the Penner and Fock coordinates are usually discussed in the context of Teichm\"uller theory, we will assume here that the construction can be analytically continued and generalized to the moduli space of flat $SL(2,\mathbb{C})$ connections and applied to the classical moduli space of 3D Schwarzschild-de Sitter. The main modification relative to the $SL(2,\mathbb{R})$ setting is that the deformation parameter $q$ is a real number instead of a phase. 

To set up the Penner-Fock construction to our setting, we first mercator project the sphere to the cylinder, so that the north and south pode become circles, enabling us to introduce two auxiliary holonomy variables $L_N$ and $L_S$. We mark a point  on both polar circles, each at identical longitude, and define $L_Z$ as the north-south holonomy between the two marked points and $L_Y$ and $L_{\tilde Y}$ as the holonomy with one winding around the cylinder. To define a triangulation of the cylinder, we choose the three holonomies $L_Z,  L_N, L_S$ and either $L_Y$ or $L_{\tilde Y}$ as our Penner coordinates, as indicated in figure 3. 
The two triangulations are related via a so-called flip move.
\begin{figure}[t]
\centering
\vspace{-2mm}
\begin{tikzpicture}[scale=.48]
\draw[thick] (0,0) -- (4,0) -- (4,4) -- (0,4) -- (0,0);
\draw[thick,cyan!60!blue] (0,4) -- (4,0);
\draw[thin,fill] (0,0) circle (.6mm);
\draw[thin,fill] (0,4) circle (.6mm);
\draw[thin,fill] (4,4) circle (.6mm);
\draw[thin,fill] (4,0) circle (.6mm);
\node[cyan!60!blue] at (2.4,2.4) {\footnotesize $L_Y$};
\node at (2,-.5) {\footnotesize $L_S$};
\node at (2,4.5) {\footnotesize $L_N$};
\node at (-.7,2) {\footnotesize $L_Z$};
\node at (4.7,2) {\footnotesize $L_Z$};
\end{tikzpicture}~~~~~~~~~~~~~
\begin{tikzpicture}[scale=.48]
\draw[thick] (0,0) -- (4,0) -- (4,4) -- (0,4) -- (0,0);
\draw[thick,green!60!black] (0,0) -- (4,4);
\draw[thin,fill] (0,0) circle (.6mm);
\draw[thin,fill] (0,4) circle (.6mm);
\draw[thin,fill] (4,4) circle (.6mm);
\draw[thin,fill] (4,0) circle (.6mm);
\node[green!60!black] at (1.6,2.4) {\footnotesize $L_{\tilde{Y}}$};
\node at (2,-.5) {\footnotesize $L_S$};
\node at (2,4.5) {\footnotesize $L_N$};
\node at (-.7,2) {\footnotesize $L_Z$};
\node at (4.7,2) {\footnotesize $L_Z$};
\end{tikzpicture}~~~~~~~~~~~~~~\begin{tikzpicture}[scale=.48]
\draw[thick] (0,0) -- (4,0) -- (4,4) -- (0,4) -- (0,0);
\draw[thick,blue] (0,2.1) -- (4,2.1);
\draw[thin,fill] (0,0) circle (.6mm);
\draw[thin,fill] (0,4) circle (.6mm);
\draw[thin,fill] (4,4) circle (.6mm);
\draw[thin,fill] (4,0) circle (.6mm);
\node[blue] at (2,1.5) {\footnotesize $L_A$};
\node at (2,-.5) {\footnotesize $L_S$};
\node at (2,4.5) {\footnotesize $L_N$};
\node at (-.7,2) {\footnotesize $L_Z$};
\node at (4.7,2) {\footnotesize $L_Z$};
\end{tikzpicture}
\vspace{-2mm}
\caption{The two triangulations of the two-punctured sphere (left and middle) with the Penner coordinates indicated. The $A$-cycle holonomy $L_A$ is depicted on the right. }
\vspace{-1mm}
\end{figure}
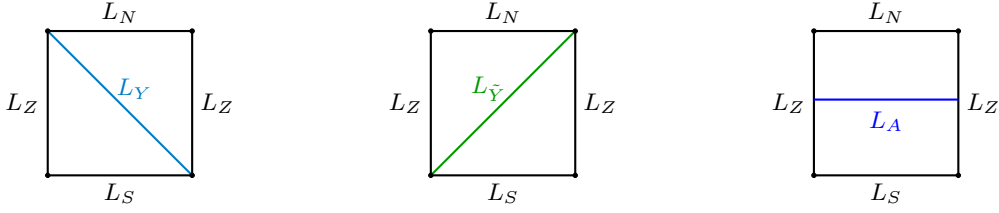

The Fock coordinates  $(Z,Y)$ or $(Z, \tilde{Y})$ associated with the two triangulations are given by are the respective cross ratios of the hyperbolic lengths of the four edges via\footnote{The Fock coordinates are usually associated with edges of the dual triangulation to the one shown in figure 3, defined by the geodesic lengths. We will not need this formulation here.}
\bea
Z^2 \spc = \spc \frac{L_ZL_Z}{L_NL_S} \qquad \qquad  Y^2\! \is\! \frac{L_Y L_Y}{L_NL_S} \qquad \qquad \widetilde{\!Y\nspc}{\spc}^2 \spc = \spc \frac{L_{\tilde{Y}}
 L_{\tilde{Y}}\mbox{\footnotesize $\strut$}} 
{L_NL_S}
\eea
We can think of these variables as regulated geodesic lengths. 

The skein relation \eqref{skeint} in Fock coordinates  reads
\bea
\label{skeinthree}
 Z L_A \is q^{\frac 1 2} \spc Y\, +\, q^{-{\frac 1 2}}\spc {\,\widetilde{\!Y\nspc}} 
\eea
The holonomy $L_A$ around the A-cycle is not a Fock coordinate but can be expressed in terms of them. The remaining data we need are (i) the commutation relation between the Fock coordinates and
 (ii) a relation between $\tilde{Y}$ in terms of $Y$ and~$Z$. The skein relation \eqref{skeinthree} will then give us the sought-after exact quantum expression for~$L_A$. Both step (i) and step (ii) are well understood. 
 
The commutation relations between the Fock variables take the form of a Heisenberg algebra
\bea
\label{heisone}
Z\spc Y \! \is \!  q\spc  Y\nspc Z, \quad \  \quad Z\spc \tilde{Y}\spc =\spc q^{-1}\spc \tilde{Y}\nspc Z 
\eea
These operator relations can be straightforwardly derived from the Chern-Simons commutators \eqref{acom} or from the local skein relation \eqref{skein}. We will not elaborate this derivation here except to note that it implies the dual form  $L_A Z = q^{-\frac 1 2} \spc Y\, +\, q^{{\frac 1 2}}\spc {\,\widetilde{\!Y\nspc}}$
of the skein relation \eqref{skeinthree}. The simplicity of these commutation relations are a key property of the Fock coordinates that  underscores their usefulness for setting up the quantum theory. 

The relationship between the coordinate $Y$ and the flipped coordinate $\tilde{Y}$ follows from the following special relation between the hyperbolic lengths of the quadrilateral with consecutive lengths $L_Z, L_N, L_Z, L_S$ and diagonals $L_Y$ and $L_{\tilde Y}$
\bea
\label{ptolemy}
\qquad \qquad \qquad \qquad\qquad L\raisebox{1.5pt}{${}_{\tilde{Y}}$}L_{Y} \! \is \! L_NL_S + L_Z L_Z\qquad\qquad\qquad\ \ \ \raisebox{0mm}{$\Bigl(\spc\raisebox{-6pt}{$\stackrel{\mbox{\small Ptolemy\tiny$\strut$}}{\mbox{\small theorem$\strut$}}$}\spc \Bigr)$} 
\eea
or in pictures \cite{roger2014skein}\\[-7mm]
\bea
\qquad \raisebox{-7mm}{$\begin{tikzpicture}[scale=.35]
\draw[dotted] (0,0) -- (4,0) -- (4,4) -- (0,4) -- (0,0);
\draw[thick,cyan!30!black] (0,4) -- (4,0);
\draw[thick,green!30!black] (0,0) -- (4,4);
\node[cyan!30!black] at (.75,2.5) {\tiny $Y$};
\node[green!30!black] at (3.25,2.5) {\tiny $\tilde{Y}$};
\end{tikzpicture}~~~\raisebox{6mm}{$=$}~~~\begin{tikzpicture}[scale=.35]
\draw[dotted] (0,0) -- (4,0) -- (4,4) -- (0,4) -- (0,0);
\draw[dotted] (0,4) -- (4,0);
\draw[dotted] (0,0) -- (4,4);
\draw[thick] (0,4) -- (4,4);
\draw[thick] (0,0) -- (4,0);
\node at (2,3.4) {\tiny $N$};
\node at (2,0.6) {\tiny $S$};
\end{tikzpicture}~~~\raisebox{6mm}{$+$}~~~\begin{tikzpicture}[scale=.35]
\draw[dotted] (0,0) -- (4,0) -- (4,4) -- (0,4) -- (0,0);
\draw[dotted] (0,4) -- (4,0);
\draw[dotted] (0,0) -- (4,4);
\draw[thick] (0,4) -- (0,0);
\draw[thick] (4,0) -- (4,4);
\node at (.6,2) {\tiny $Z$};
\node at (3.4,2) {\tiny $Z$};
\end{tikzpicture}$}\\[-2mm]\nonumber
\eea
This relation can be recognized as the hyperbolic generalization of the classic theorem of Ptolemy in euclidean geometry, relating the product of the lengths of the diagonals of a quadrilateral to the sum of the product of the lengths of the opposite edges.\footnote{Ptolemy's theorem states that a euclidean quadrilateral with consecutive side-lengths $\ell_A, \ell_B, \ell_C, \ell_D$ and diagonal lengths $\ell_E, \ell_F$ inscribes in a circle if and only if $\ell_E\ell_F= \ell_A\ell_C + \ell_B\ell_D$. The proof of the relation~\eqref{ptolemy} in the hyperbolic setting is given in \cite{penner1987decorated}. Note that in our case, the four corner points project to the north and south pode. The corner points thus lie on a single geodesic and two of the sides have the same length. One can also derive the Ptolemy relation \eqref{ptolemy}-\eqref{ptolemyt} by directly applying the local skein relation \eqref{skein} to $Y\tilde{Y}$ or $\tilde{Y}Y$.} In terms of the Fock coordinates, the Ptolemy relation reads
\bea
\label{ptolemyt}
\tilde{Y}\smpc {Y} \!\is \! 1 + q\spc Z^2, \spc \quad\ \quad Y\smpc \tilde{Y} \spc = \spc 1+ q^{-1} Z^2
\eea
The above relations are consistent with the commutation relations \eqref{heisone}.  Multiplying the skein relation \eqref{skeinthree} with $Z^{-1}$ 
we obtain our final expression for the A-cycle holonomy 
\bea
\label{lala}
L_A \!\is \! q^{\frac 1 2} \spc Z^{-1} Y  
+ q^{{\frac 1 2}}\, \tilde{Y}Z^{-1} \small {}_{\strut}^{\strut} 
\eea
This formula for the A-cycle holonomy, together with the Ptolemy relation \eqref{ptolemyt}  and commutation relations \eqref{heisone}, is the main result of this subsection.

As a side comment, we note that it is often conventional to define small caps Fock coordinates 
\bea
(Z,Y,\tilde{Y}) \!\is\! (e^{z}, e^{y},e^{\tilde{y}}).
\eea
The small coordinates satisfy canonical commutation relations and a logarthmic Ptolemy theorem 
\bea
[\smpc z,y\spc]\spc  =  -[\smpc z,\tilde{y}\spc]  \!\is\! \frac{2\pi}{\kappa},\qquad \qquad
\tilde{y}  =-y + \log(1+e^{2z})_{\strut}.
\eea
Note that  there's no factor of $i$ in the right-hand side of the canonical commutators, so the $\hbar$ parameter is imaginary.
The A-cycle holonomy operator in these coordinates takes the form
\bea
\label{lalas}
L_A \! \is \! 2 \cosh(y-z) + e^{-{z-y}}
\eea
This formula for the hyperbolic length $L_A$ is a classical relation between geodesic lengths. Its quantum version is well known to experts in quantum Teichm\"uller theory \cite{Chekhov:1999tn, Teschner:2003em,Terashima:2011qi}. In that case, the deformation parameter $q$ is a phase. In the next subsection, we will determine the eigen spectrum of $L_A$ for the case that $q$ is a real number between 0 and 1.\footnote{ The same formula \eqref{lalas} also appeared in recent work on the reformulation of the double scaled SYK Hamiltonian in terms of q-deformed Schwarzian quantum mechanics \cite{Blommaert:2023opb}}

\subsection{Spectrum}

\vspace{-1.5mm}

We are now in a position to compare our Hamiltonian with the double scaled SYK model. We will find that the two precisely match. We then use this match to compute the energy spectrum and partition function of the de Sitter gravity model.

\def\mfp{{\smpc\mathfrak{p}}}
\def\mfn{{\smpc\mathfrak{n}}}

\def\mfns{\mbox{\footnotesize $\smpc\mathfrak{n}$}}

It is convenient to introduce slightly modified canonical phase space coordinates $\mfp$ and $\mfn$ via
\bea
e^{i\mfp} \!\is \! \mfq^{\frac 1 4} Z^{-1} Y\qquad \qquad \mfq^{\smpc\mathfrak{n}}\nspc =-\spc \mfq^{\frac 1 2 } Z^{-2},
\eea
with $ \mfq \equiv q^2$. The commutator algebra between the canonical variables reads
\bea
\label{pncom}
  [\smpc \mfn, e^ {i \mfp}]\!\is \! e^{i\mfp}  \quad \ \ \ \ \quad  [\mfn, e^ {-i \mfp}]\spc = \spc- e^{- i\mfp}.
\eea
The A-cycle holonomy operator then takes the form
\bea
\label{lalat}
 L_A \!\is\! 2\spc {\cos \mfp -e^{-i\mfp}\, \nspc\mfq^{\mfns} }
\eea
We see that $\mfp$ appears in the above relations only via the combination $e^{\pm i\mfp}$. It is therefore natural to identify $\mfp$ as an angular variable defined modulo $2\pi$ and view the dual coordinate $\mfn$ as having an integer spectrum. Note that both conclusions crucially depend on the fact that $\hbar$ in \eqref{acom} and the CS level $k=i\kappa$ are imaginary, so that the deformation parameter $\mfq = e^{-\frac{4\pi}{\kappa}}$ is real.

We now introduce the operators $\mfa$, $\mfa^\dag$ via
\bea
\label{adeft}
\mfa^\dag \spc = \spc \, \frac{e^{i\mfp}}{\!\!\sqrt{1\nspc-\nspc\mfq}\,}, \quad & &\quad \mfa \spc = \spc \spc e^{-i\mfp}\, \frac{1\nspc -\nspc\mfq^{\mfns}\!}{\!\!\sqrt{1\nspc -\nspc\mfq}}{}_{\strut}
\eea
 From the commutation relations \eqref{pncom}, one immediately derives that the operators  $\mfa$ and $\mfa^\dag$ satisfy the $\mfq$-deformed oscillator algebra 
\bea
\mfa \spc \mfa^\dag \spc = \spc \frac{1-\mfq^{\mfn + 1}\!\!}{{1\nspc -\nspc\mfq}}, \quad & & \quad \mfa^\dag\smpc \mfa \spc = \spc \frac{1- \mfq^{\mfn} }{{1\nspc -\nspc\mfq}}, \qquad \quad
[\mfa,\mfa^\dag]_\mfq \spc = \spc 1.
\eea
Here the $\mfq$-commutator is defined as $
[A,B]_\mfq\equiv  AB - \mfq BA.$
The relation between the $\mfq$-oscillators and the Fock variables of the 3D de Sitter gravity reads
\bea
\label{adef}
{\sqrt{{1\! -\!\mfq}}}  \; \mfa^\dag \spc = \spc {\mfq^{\frac 1 4} Z^{-1} Y}, \ \ & & \ \  {\sqrt{{1\! -\!\mfq}}} \; \mfa \spc = \spc {\mfq^{\frac 1 4}  \tilde{Y}Z^{-1}}
\eea
We see that the two oscillators $\mfa^\dag$ and $\mfa$ are interchanged under the parity transformation that exchanges the two diagonal holonomies $Y$ and $\tilde{Y}$.

The $\mfq$-deformed oscillator algebra plays a central role in the solution of double scaled SYK model. This correspondence directs us towards defining the Hamiltonian of 3D de Sitter gravity in terms of the A-cycle holonomy operator $L_A$ via
\bea
\HH  \!\is \!   \frac{L_A}{\sqrt{{1\nspc -\nspc\mfq}}} \spc = \spc \mfa^\dag + \mfa
\eea
Here we used the expressions \eqref{lala}-\eqref{lalat} for $L_A$ and the definitions \eqref{adeft}-\eqref{adef} of the $\mfq$-oscillators. This identification of the A-cycle holonomy with the Hamiltonian is well-motivated from the gravity perspective given that $L_A$ measures the deficit angle, and thus the energy, of the Schwarzschild-de Sitter spacetime.

As discussed in section 2.1, the Fock coordinate $z$ is, up to a shift by $i\pi$ arising from the integral from $0$ to $\pi$ of the $\rho$ component of the gauge connection \eqref{apm}, identified with the time-difference between the clocks of two observers located at the the north and south podes. The relation 
\bea
z \! \is \! i\pi + \mbox{\Large $\frac{2\pi}{\kappa}$}\, \mfn
\eea
shows that this time difference becomes discrete in the quantum theory at a time scale of order ${4\pi}/{\kappa} \simeq  16\pi G_n \simeq 1-\mfq$. This conclusion mirrors ideas by ’t Hooft and others that time evolution in 3D quantum gravity with non-negative cosmological constant should be thought of as being discrete \cite{tHooft:1996ziz}. The reasoning is that, since massive states in 3D gravity with $\Lambda>0$ create conical defects, the total energy is bounded. Hence time differences can be measured with finite accuracy. 
Moreover, the fact that angles are naturally periodic suggests that the dual variable has a discrete spectrum. Here we have reached the same conclusion via the first order formulation of 3D gravity.

We can construct an eigenbasis of the discrete time variable  $\mfn$ normalized such that
\bea
\mfa^\dag | n\ra = | n+1\ra, \qquad \quad \mfa | n\ra\spc =\spc [n]_\mfq |n-1\ra, \qquad 
\mfn\, | \smpc n \smpc \ra \! \is \! n \, | \smpc n \smpc \ra.
\eea
The Hamiltonian $H$ acts on this eigenbasis as follows
\bea
\label{hrecur}
\qquad \qquad\ \HH|\spc n\spc \ra \! \is \! |\smpc n\nspc +\nspc 1\ra +  [n]_\mfq\; |\smpc n\nspc-\nspc 1\ra,\qquad \ \  [n]_\mfq\spc\equiv\spc \frac{1-\mfq^{n}\!}{1-\mfq} 
\eea
This equation takes the exact same form as the action \eqref{hamrule} of the Hamiltonian of double scaled SYK on the chord number eigen basis. We view this match as strong evidence for a holographic DSSYK-de Sitter correspondence.

From here on we can just follow the exact same steps that were used to compute the double scaled SYK spectrum. The only assumption we need to make is that the Hamiltonian $\HH$ defines a hermitian operator on the physical Hilbert space of the de Sitter gravity theory. Adopting the standard DSSYK  notation, we parametrize the eigen states and eigen values of the Hamiltonian in terms of a spectral angle via 
\bea
\label{eigenh}
\HH |\spc \theta\spc \ra \spc = \spc \frac{2 \cos \theta}{\sqrt{1-\mfq}} \spc |\spc \theta \spc \ra 
\eea
Equations \eqref{hrecur} and \eqref{eigenh} then imply the recursive formula
\bea\qquad   \la  \spc \theta\spc|\smpc n \!+\!1\smpc \ra+[n]_\mfq\spc \la\spc \theta\spc  | \smpc n\!-\!1 \ra\! \is \! 
x  \la\spc \theta\spc |\smpc n \smpc \ra, \qquad \ x\equiv \small \frac{2 \cos \theta}{\sqrt{1-\mfq}}\eea
which is solved by $\la \spc \theta \spc | \smpc n\smpc  \ra =
\, H_n(x| \mfq)$
where $H_n(x| \mfq)$ denotes the $n$-th $\mfq$-Hermite polynomial. The $\mfq$-Hermite polynomials are defined via recursive relation $
 H_{n+1}(x|\mfq) + [n]_\mfq H_{n-1}(x|\mfq)  = x \, H_n(x,\mfq),$
 with initial condition  $H_0(x|\mfq) = 1$, $H_1(x|\mfq) =x$. They form an orthogonal basis of functions with respect to the inner product defined by the integral over $\theta$ 
\bea
\la \smpc n \smpc |\smpc m \smpc \ra \spc = \spc \int_0^\pi \!\! d\theta \, \rho(\theta) H_n(x|q) H_m(x|q)\!\is\! \delta_{nm} \spc [n]_{\mfq}!
\eea
with integration measure\\[-8mm]
\bea
\rho(\theta) \! \is\! (e^{\pm 2i\theta};q)_\infty.
\eea
The energy eigenstates $|\theta\ra$ thus form a delta-function normalized basis with spectral density $\rho(\theta)$
\bea
\la \spc \theta_1 \spc | \spc \theta_2 \spc \ra \! \is \! \sum_n \frac{1}{[n]_q!\!} \; H_n(x_1|q) H_n(x_2|q) \spc =\spc \frac{\delta(\theta_1\nspc-\theta_2)}{\rho(\theta_1)}
\eea
This formula tells us that the spectral density of energy eigen states obtained by quantizing the Schwarzschild-de Sitter spacetime is given by $\rho(\theta)$.

\def\sfs{{\sf s}}
\def\sfr{{\sf r}}

\def\x0{\mbox{\tiny $x_0$}}
\def\xo{\mbox{\tiny $x_1$}}

\subsection{Partition function}

\vspace{-1.5mm}

We have obtained the spectrum of pure 3D de Sitter gravity by quantization of the moduli space of non-rotating Schwarzschild-de Sitter spacetimes and found that it matches with the spectral density of DSSYK.  Using this result, and our identification of the de Sitter Hamiltonian with the A-cycle holonomy, we can formally define the thermal partition function of de Sitter gravity via~\cite{Blommaert:2023opb}
\bea
Z(\beta) \spc = \la 0 | e^{-\beta H} | 0 \ra 
 \is \!\int\! dE(\theta) \spc \rho_E(\theta) \spc e^{-\beta E(\theta)}\\[-9mm]\nonumber
 \eea
 with\\[-7mm]
 \bea
 E(\theta) =  -\frac{2\cos\theta}{\sqrt{\lambda(1\nspc -\nspc\mfq)}} \quad & &\quad \rho_E(\theta) =  e^{S_0} \vartheta_1(2\theta,\mfq)\,.
 \eea 
 Here we restored a factor of $\lambda^{-1/2}$ in the definition of the Hamiltonian and energy and included an overall constant pre-factor $e^{S_0}$ in the spectral density.  Via the identification of the angle $\theta$ with the deficit angle, we can interpret this partition function as a thermal ensemble of non-rotating Schwarzschild-de Sitter spacetimes. As discussed above, the condition that the spacetime is non-rotating is implemented by requiring that the A-cycle holonomies $L_A^+$ and $L_A^-$ are identical. 

Though we have formulated our model in terms of 3D variables, it is relevant to note that, since we restrict the phase space to non-rotating spacetimes, the quantum system defined by its quantization can be viewed as a 2D JT-like gravity theory obtained via an s-wave reduction of 3D Einstein-de Sitter gravity. As seen from the form \eqref{statmetric}-\eqref{hopfm} of the static metric, the dimensionally reduced metric looks like that of 2D de Sitter space.  A spatial slice of the 2D spacetime forms line that stretches between the north and south pode, while the A-cycle holonomy $L_A$ descends to the extremal value of the 2D dilaton field on this slice. 
Our 3D quantum Hamiltonian obtained above indeed looks like a $\mfq$-deformed Hamiltonian of 2D JT gravity \cite{Harlow:2018tqv}\cite{Blommaert:2023opb}.

The spectral density $\rho_E(\theta)$ takes the form of a $\mfq$-deformed gaussian distribution. This is made most manifest by using the modular transformation property of the theta function to write it as  
\bea
\label{rhothetasum}
\rho_E(\theta)  \! \is \! e^{S_0} 
\; \sum_{n} (-1)^n e^{-\frac{1}{\! 4\pi G_N\!}\spc (\theta + \pi (n-\frac 1 2))^2} 
\eea
Here we used $\mfq = e^{-\frac{4\pi}{\kappa}} = e^{-8\pi G_N}$, so the modular transformed nome equals $\tilde{\mfq} = e^{-\pi \kappa} = e^{-\frac{\pi}{2G_{\! N}\!}}$. We see that, just like the energy $E(\theta)$ itself, the spectral density is a periodic function of $\theta$ and takes the form of a sum of gaussians with maxima at
\bea
2\theta_n \! \is \! (1-2n) \pi.
\eea
The maxima occur at the location  where the energy vanishes, $E(\theta_n) = 0$.

Comparing with the geometric definition of the A-cycle holonomy, we are led to identify the DSSYK spectral angle with the Schwarzschild-de Sitter deficit angle via
\bea
\label{defid}
2\pi \alpha \! \is \! 2\pi - 2(\theta_0 - \theta) \qquad \qquad \theta_0 \equiv \frac \pi 2
\eea
Via this identification, it is natural to interpret the sum \eqref{rhothetasum} as sum over semiclassical saddle geometries with deficit angle (or excess) equal to
\bea
2\pi \alpha_n \!\is \! 2\pi - 2(\theta_0 - \theta_n)\spc = \spc  2\pi (1- n)
\eea
As mentioned earlier, the first order formulation of 3D de Sitter gravity allows for SdS configurations with deficit angles that exceed $2\pi$ or geometries with angle excess  $-2\pi \alpha>0$. 

Given that the spectral density \eqref{rhothetasum} is periodic in the spectral angle $\theta$, it is natural to restrict $\theta$ to the interval from 0 to $\pi$. The expression \eqref{rhothetasum} then indicates that the state labeled by $\theta$ should be thought of as the superposition of semi-classical geometries with unfolded deficit angle $2\pi \alpha = 2(\theta_0 -  \theta) + 2\pi(1- n)$ with $n \in \mathbb{Z}$. This interpretation of \eqref{rhothetasum} as a sum over semi-classical  configurations related via large gauge transformations is further supported by the fact that the pre-factor $\frac{1}{16\pi G_N}$ in the exponents matches with the pre-coefficient of the first order Einstein action \eqref{csaction}. It would be important to verify this by means of an exact evaluation of the de Sitter gravity functional integral along the lines of \cite{Castro:2011xb,Hikida:2021ese,Hikida:2022ltr}. We leave this check for future work.

From \eqref{rhothetasum} we see that the entropy $S(\theta) = \log \rho_E(\theta)$ in the semi-classical $G_N\to 0$ limit reads
\bea
\label{szero}
S(\theta) \! \is \! S_0 -\frac{1}{\! 4\pi G_N\!}\spc \bigl(\theta -\theta_0\bigr)^2.
\eea
$S(\theta)$ has a maximum at the special angle $\theta_0 = \frac \pi 2$. The energy and temperature near $\theta_0$ behave as
\bea
\label{etzero}
E(\theta) \! & \!\simeq \!& \! \frac{\theta-\theta_0}{4\pi G_N}, \qquad\qquad \beta\spc = \spc \frac{dS}{dE}\, \simeq \, 2(\theta_0 - \theta).
\eea
We would like to identify the corresponding maximal entropy state $|\theta_0\ra$ with pure de Sitter space\footnote{
More generally, we could identify the quantum state of the de Sitter spacetime as the micro-canonical  state $$\rho_{\rm dS} = \frac{1}{N} \sum_{E} |E\ra\la E|,$$ where the sum runs over $N$ energy eigen states with energy very close to $E_0\equiv E(\theta_0) = 0$.}
\bea
|\Psi_{\rm dS}\ra \! \is \! |\theta_0 \ra. 
\eea
There are indeed many indications that pure de Sitter spacetime  describes a maximal entropy state at infinite temperature. On the other hand, this characterization looks to be at odds with the physical observations of a static observer that the de Sitter spacetime should have a finite temperature equal to $T_{\rm dS} = \frac{1}{2\pi}$ in de Sitter units. In the next subsection, we describe how the two perspectives can be reconciled following the earlier work \cite{ustwo} relating DSSYK and de Sitter two-point functions. Here we point to another possible explanation for why de Sitter spacetime seems to have more than one definition of temperature based on the proposed interpretation of the spectral density \eqref{rhothetasum} as a sum of semi-classical saddle points. 

As above, we consider the regime where $\theta$ approaches the infinite temperature point $\theta_0 = \frac{\pi}{2}$. The identifications \eqref{szero} and \eqref{etzero} of the entropy and temperature are based on the assumption that all saddle points contribute and that the leading contribution is therefore coming from the $n=0$ saddle point. Suppose, however, that for some special correlation functions this leading saddle point does not contribute, but the next to leading saddle with $n=1$ does. Let us compute the entropy and inverse temperature for this $n=1$ saddle point. We find
\bea
S_1(\theta) \! \is \! S_0 -\frac{1}{\! 4\pi G_N\!}\spc \bigl(\theta -\theta_0-\pi\bigr)^2, \\[2mm]
T_1 \spc=\spc\frac{dE}{dS_1} \!&\! \simeq \!&\!\frac{1}{2\pi + 2(\theta_0-\theta)} \spc \simeq \spc
\spc \frac{1 - 4 G_N E}{2\pi}.
\eea
This formula for the temperature should be compared with the known relation between the energy and temperature of a 3D Schwarzschild-de Sitter spacetime
\bea
T_{\rm SdS} \!\is \! \frac{\sqrt{1 - 8 G_N E}}{2\pi}
\eea
This match supports our proposal that the different terms in the spectral density $\rho_E(\theta)$ correspond to different classical geometries, each with different entropies and temperatures, and that pure de Sitter spacetime corresponds to the first sub-leading saddle point with $n=1$.  Note that the first equation in \eqref{etzero} leads to the correct identification $2\pi \alpha = 2(\theta_0-\theta) = 8 \pi G_N E$ for the deficit angle.

\subsection{Two-point function}

As a final exercise, we now outline how one can use the above results to compute the exact quantum expression for the scalar two-point function in 3D de Sitter quantum gravity. Schematically, 
\bea
\label{twopt}
\tilde{G}_\Delta(\tau) \! \is \! \bigl\la \tilde{\phi}_\Delta(x_1) \phi_\Delta(x_0)\bigr\ra_{\rm dS}\, = \, \bigl\la\Psi_{\rm dS}|{\cal W}_{\Delta}(\tau)
|\Psi_{\rm dS} \bigr\ra 
\eea
where ${\cal W}_\Delta (\tau) 
 =  e{\raisebox{5pt}{\footnotesize $i \int_{\x0}^{\xo}m\spc ds$}}$ denotes the wordline action of a massive scalar particle of mass $m$ that travels between the points $x_1$ and $x_2$. For concreteness, we take $x_1$ to be localized at the south pode and $x_2$ at the north pode. The physical expectation is that the scalar two-point function \eqref{twopt} will only depend in this geodesic distance 
$\tau = \tau(x_1,x_2)$ between the two points. We can equivalently represent ${\cal W}_\Delta (\tau)$ as a gravitational Wilson line operator $\,{\cal W}_{\Delta}  = 
P e\,{\raisebox{4.5pt}{${\int_{\x0}^{\xo} \!\mbox{\footnotesize$\!{A_-}$}}$}}\, P e\,\raisebox{4.5pt}{$\int_{\x0}^{\xo} \! \mbox{\footnotesize $\!{A_+}$}$}$ with second casimir $c_2(\Delta) =  \Delta(1\nspc -\nspc \Delta)  = {m_\ddelta^2}/{4}$
and end-points anchored  via a suitable $SL(2,\mathbb{C})$ invariant tensor to the observer worldlines at the podes.  

\addtolength{\parskip}{-1mm}
In a moment, we will write an expression for ${\cal W}_{\Delta}(\tau)$ in terms of the north-south holonomy operator $Z$ introduced in section 2.5, or more practically, the operator $\mfq^{\mfns}$ introduced in section~2.6. This expression will be informed by four physical data points:

\addtolength{\baselineskip}{-1mm}
\begin{enumerate}
\addtolength{\baselineskip}{-1mm}
\item{the above description of ${\cal W}_{\Delta}(\tau)$ as a worldline operator or gravitational Wilson line, }
 \item{ the expression for the DSSYK two-point functions in terms of the $\mfq$-oscillators}
\item{the requirement that in the semi-classical limit, it should reproduce the known expression 
\bea
\label{twopt}
\qquad \tilde{G}_\Delta(\tau)_{|\mfq\to 1}  \is \frac{2\sinh\mu\tau}{\pi \sinh \tau}
\qquad \qquad
\mu \equiv 2\Delta -1 = \sqrt{1-m^2}
\eea
for the anti-podal de Sitter two-point function of a massive scalar of mass $m$, and}
 \item{the correspondence between two-point functions in DSSYK and complex Liouville CFT \cite{ustwotwo}}
\addtolength{\baselineskip}{1mm}
\end{enumerate}
\vspace{-1.5mm}

\addtolength{\baselineskip}{1mm}
\addtolength{\parskip}{1mm}
Guided by the above input data, we propose that the gravitational Wilson line is given by the following expression (c.f. \cite{Lin:2023trc}\cite{Okuyama:2024yya})
\bea
\label{wdelta}
{\cal W}_{\Delta}(\tau) \is  \mfq^{\mfns(1\nspc-\nspc \Delta)} e^{-i\tau {H}} \mfq^{\mfns  \Delta} \,=\, Z^{-2(1-\Delta)} e^{-i\tau {H}} Z^{-2\Delta} 
\eea
The second expression on the right reflects the fact that the particle worldine defines a holonomy variable between the north and south pode. Schematically one can think of the factor $Z^{-2\Delta}$ as the Wilson line of the $A_+$ gauge field and $Z^{-2(1-\Delta)}$  as the Wilson line of the $A_-$ gauge field. The combination of the two Wilson lines must be constructed to satisfy the physical constraint that it is free of framing anomalies \cite{Witten:1988hc}. This is requirement is analogous to the condition that physical vertex operators in string theory must be diffeomorphism invariant. As we will see shortly, the operator $e^{-i \tau H}$ controls the  geodesic time-difference between the two end-points of the worldline. 

The operators $\mfq^{\mfns  \Delta}$ and $\mfq^{\mfns(1\nspc-\nspc \Delta)}$ look familiar to experts in double scaled SYK: the matrix elements of these operators between energy eigen states represent the two-point function ${\cal O}_\Delta$ and ${\cal O}_{1-\Delta}$, respectively \cite{Berkooz:2018jqr,Lin:2023trc,Okuyama:2024yya}.  The explicit form of the matrix elements are given by
\bea
\la \spc \theta_1 \spc | \mfq^{\spc\mfns\Delta} | \spc \theta_2 \spc \ra \! \is \! \sum_n \frac{\mfq^{n\Delta}}{[n]_q!} \, H_n(x_1|\mfq) H_n(x_2|\mfq) \, = \, \frac{(\mfq^{2\Delta}; \mfq)_\infty\!}{(\mfq^\Delta e^{\pm i \theta_1\pm i \theta_2} ; \mfq)_\infty\!}
\eea
The formula \eqref{wdelta} for the gravitational Wilson line produces via \eqref{twopt} the two-point function of two physical operators in the doubled DSSYK model introduced and studied in \cite{ustwo}.
Using the known relation between Pochhammer symbols and Jacobi theta functions, we compute
\bea 
\label{gexact}
\tilde{G}_\Delta(\tau) \is \! \la\Psi_{\rm dS}|{\cal W}_{\Delta}(\tau)
|\Psi_{\rm dS} \ra 
 \spc = \spc \la\smpc\theta_0| \mfq^{\mfns(1\nspc-\nspc \Delta)} e^{-i\tau {H}} \mfq^{\mfns  \Delta}|\smpc \theta_0\ra \nonumber\\[3.5mm]
\! \is \!\, \int\!\! d\theta_1\rho(\theta_1) \spc e^{- \tau E(\theta_1)}\!\, 
 \la \smpc \theta_0 |  \mfq^{\mfns(1\nspc-\nspc \Delta)} | \smpc \theta_1\ra \la \smpc \theta_1  |  \mfq^{\mfns\Delta} | \smpc \theta_0 \ra
\\[2mm]\nonumber
 \! \is \!\int\! dE(\theta_1)\, e^{-\tau E(\theta_1)} \; \frac{\vartheta_1  \bigl( 2\theta_{1},\mfq \bigr) \vartheta_1  \bigl( 2\hbar\Delta, \mfq \bigr) \strut}
{\smpc \vartheta_1  \bigl(\hbar\Delta \pm  \theta_0\pm \theta_1,\mfq \bigr) \strut} .
\eea
It was shown in \cite{ustwo} that this final expression for the two-point function reduces in the $\mfq\to 1$ limit to the semi-classical two-point function of a massive scalar field in 3D de Sitter space in terms of the proper time difference $\tau$. This match further motivates our proposed expression \eqref{wdelta} for the matter Wilson line. This semi-classical correspondence should help inform future investigations of the semi-classical limit of 3D de Sitter quantum gravity and its relationship with DSSYK.

\section{Concluding remarks}
\vspace{-1.5mm}

In this paper, we performed a new quantitative check of the correspondence  between 3D de Sitter quantum gravity and the double scaled SYK model by establishing a direct match between the gravity Hamiltonian and the combinatorial rules that guide the exact computation of the DSSYK spectrum and correlation functions. The holographic dictionary equates the SYK Hamiltonian and matter chords to gravitational Wilson lines that respectively measure the conical deficit of the Schwarzschild-de Sitter spacetime and describe the world line action of matter particles propagating between the north and south pode. The fact that the found link ties together elementary quantities on both sides is an encouraging sign that the identifications we have found are just a small corner of a more complete correspondence. Given that both sides of the duality are exactly soluble systems, there are more directions to explore and detailed tests to perform. We mention a few below.

It may sound like a bold step to directly link a 1D quantum many body system to a 3D quantum gravity system, since it would need to rely on a mechanism that produces two extra emergent dimensions. Our proposed duality is not as radical as it sounds, however. First, as argued in \cite{ustwo}, the DSSYK model should be seen as living on a pair of 1D time-like trajectories of localized observers in 3D de Sitter space. Moreover, pure 3D de Sitter gravity is a topological QFT and we restrict all states to the $s$-wave sector. The DSSYK-SdS duality studied in this paper is part of a triangle of dual correspondences  labeled by {\bf a}, {\bf b} and {\bf c} in the below diagram:

$$
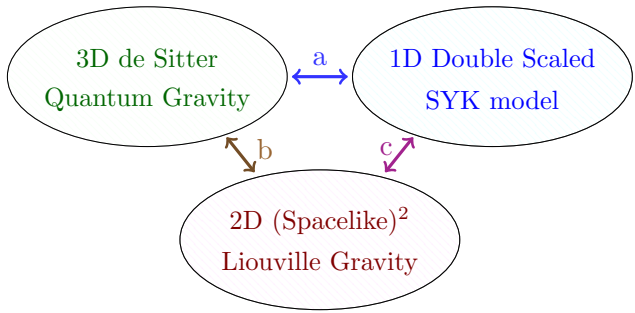
\begin{tikzpicture}[yscale=.62,xscale=.62]    
        \draw[thin, pattern={north west lines}, pattern color={cyan!20!white},opacity=.50] (3.7,3.5)  ellipse (3cm and 1.5cm);
        \draw[thin,pattern={north west lines}, pattern color={magenta!15!white},opacity=.50] (0, 0) ellipse (3cm and 1.5cm);
        \draw[thin,pattern={north west lines}, pattern color={green!25!white},opacity=.30] (-3.7, 3.5) ellipse (3cm and 1.5cm);
           \draw[blue!80!white,very thick, <->] (-.6,3.5) -- (0,3.5) node[above,blue!80!white] {a}-- (.6,3.5) ;   
           \draw[brown!60!black,very thick, <->] (-2,2.2) -- (-1.8,1.95) node[right,brown!80!black] {~b}-- (-1.4,1.45) ;
           \draw[magenta!60!blue,very thick, <->] (2,2.2) -- (1.8,1.95) node[left,magenta!60!blue] {c}-- (1.4,1.45) ;
        \draw[very thin] (3.7, 3.5) ellipse (3cm and 1.5cm);
        \draw[very thin] (-3.7, 3.5) ellipse (3cm and 1.5cm);
         \draw[very thin] (0, 0) ellipse (3cm and 1.5cm);
         \node[green!40!black] at (-3.7,3.9) {\small 3D de Sitter};
         \node[green!40!black] at (-3.7,3) {\small Quantum Gravity};     \node[red!50!black] at (0,0.4) {\small 2D (Spacelike)$^2$};
         \node[red!50!black] at (0,-.5) {\small Liouville Gravity};
         \node[blue] at (3.7,3.9) {\small 1D Double Scaled};
         \node[blue] at (3.7,3) {\small SYK model}; 
\end{tikzpicture}
$$

\noindent
Here we studied the direct 3D-1D link ${\bf a}$. The other two links are elaborated in a companion paper \cite{ustwotwo}. Both links are found to be on firm footing and give additional evidence for  link~${\bf a}$. The 3D-2D link {\bf b} follows from the well-known correspondence between 3D topological Chern-Simons theory on a three-manifold with boundary and 2D chiral conformal field theory \cite{Witten:1988hc}\cite{Verlinde:1989ua}. As explained in \cite{ustwotwo} (see also \cite{Klemm:2002ir}\cite{Collier:2023fwi},\cite{Collier:2023cyw},\cite{spacelike-lcft-string}), the same familiar analysis applied to 3D de Sitter gravity gives rise to a non-chiral pair of spacelike Liouville CFTs, each with a complex central charge of the form $c_\pm = 13\pm i\gamma$ and total central charge $c_++c_-=26$. The 3D gravitational Wilson-lines studied in this paper correspond to Verlinde loop operators in the Liouville gravity theory \cite{Gaiotto:2014lma}. The 2D-1D link  {\bf c} between DSSYK and Liouville CFT was first noted in \cite{Cotler:2016fpe}\cite{douglas-talk-kitp}. In \cite{ustwotwo}, this link is firmed up by showing that the DSSYK two-point function \eqref{gexact} exactly matches with the boundary two-point function of the spacelike Liouville gravity theory on the disk.

We end with a few comments about future directions.

\noindent
\textit{Quantum Group Symmetry.} \\[.5mm]
A prominent link between double scaled SYK, 2D Virasoro-Liouville gravity, and 3D de Sitter quantum gravity is that all three incorporate a $U_{\!\mathsf q}(\mathfrak{sl}_2)$ symmetry algebra
\bea
\label{qalg}
K EK^{-1} = q E, \qquad K F K^{-1} = q^{-1} F, \qquad [E,F\smpc]_q = \frac{K - K^{-1}}{q- q^{-1}\!}
\eea
Quantum groups are somewhat subtle and seemingly esoteric, but have a clear physical role in encoding the braiding properties of correlations functions in 2D conformal field theory and in 3D topological QFT.
In SYK, the quantum group algebra acts on the Hilbert space with matter lines by means of suitable bilinear combinations of the chord creation and annihilation operators \cite{Lin:2023trc}.  On the gravity side, the $U_{\!\mathsf q}(\mathfrak{sl}_2)$ symmetry governs the fusion and braiding interactions between the world lines of massive point particles. The construction of the quantum group generators in this setting involves considering open gravitational Wilson lines that end on defects \cite{Witten:1989rw}. In the $q\to 1$ limit, the $U_{\!\mathsf q}(\mathfrak{sl}_2)$ algebra \eqref{qalg} reduces to a classical $SU(1,1)$ Lie algebra. It is tempting to identify these transformations with isometries acting on 3D de Sitter space that preserve the $\mathbb{Z}_2$ identification \eqref{pete}. 
The deviations between the quantum and classical symmetry generators is of order $1-q \simeq \lambda$. This sets the ratio between the Planck scale and the de Sitter radius.

\smallskip

\noindent
\textit{Scattering and 6j-Symbols}\\[.5mm]
The quantum group symmetry imposes powerful constraints on correlation functions. Correlators in 3D gravity are build up from partial waves that satisfy selection rules that mirror the selection rules of the representation ring of $U_{q}(\mathfrak{sl}_2)$. This structure is particularly helpful in the study of gravitational shockwave interactions on the gravity side and of out-of-time-ordered correlators on the SYK side \cite{Maldacena:2016hyu,Berkooz:2018qkz,Mertens:2017mtv,Lam:2018pvp}. Here we only make a brief comment about this application. 

De Sitter gravity has a horizon and shockwave solutions \cite{Hotta:1992qy,Leblond:2002ns, Aalsma:2021kle,Anegawa:2023dad}, similar to an AdS black hole space-time. However, there is a main difference: unlike in AdS space, the shockwaves can open up a communication channel between space-like separated points on opposite sides of the cosmological horizon, that otherwise would require superluminal propagation \cite{Leblond:2002ns,Aalsma:2021kle}, c.f. \cite{Gao:2016bin}. 

Figure 6 shows the thought experiment. At an early time, two observers, one at the north-pole and one at the south-pole, send a light-signal towards the cosmological horizon. To preserve the $\mathbb{Z}_2$ symmetry, we assume both signals have the same energy and are sent at the same instant $\tau_1$. The two signals carry light-like Kruskal momenta 
of equal magnitude and create gravitational shockwaves along their trajectory in the form of a discontinuous shift of the conjugate light-cone coordinates. As a result, the Penrose diagram of the de Sitter space-time elongates, opening up a gap between the future and past horizons, as shown. This semi-classical gravity experiment predicts that the corresponding four-point function exhibits a singularity when the two late-time operators ${\phi}^\pm(\tau_2)$ approach the classical arrival time of the light-signal emitted by the two early operators at time $\tau_1$.

 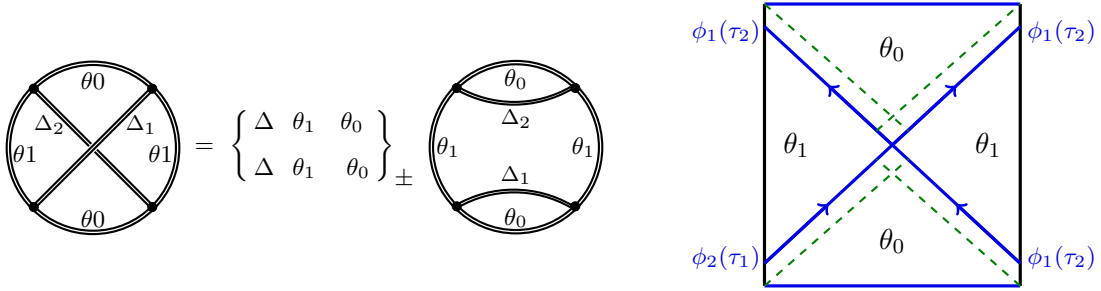
\begin{figure}
\begin{center}
\quad\raisebox{2.2cm}{\begin{tikzpicture}[scale=.75, baseline={([yshift=.35cm]current bounding box.center)}]
\draw[thick] (-1.03,1.07) -- (-.03,.07);
\draw[thick] (.03,-.07) -- (1.03,-1.07);
\draw[thick] (-1.07,1.03) -- (-.07,.03);
\draw[thick] (.07,-.03) -- (1.07,-1.03);
\draw[thick] (-1.07,-1.03) -- (1.03,1.07);
\draw[thick] (-1.03,-1.07) -- (1.07,1.03);
\draw[thick] (0,0) circle (1.47);
\draw[thick] (0,0) circle (1.53);
\draw[fill,black] (-1.05,-1.05) circle (0.08);
\draw[fill,black] (1.05,-1.05) circle (0.08);
\draw[fill,black] (-1.05,1.05) circle (0.08);
\draw[fill,black] (1.05,1.05) circle (0.08);
\draw (3.1,0) node {\footnotesize $\ \ \ \ \ \ \ \  \ = \ \left\{\begin{array}{ccc}\!\!\Delta\!&\! \theta_1\!&\! \theta_0\!\!\\[2mm]
\!\! \Delta\!&\! \theta_1\!&\ \theta_0\!\! \end{array}\right\}_{\mbox{$\pm$}}$};
\draw (1.2,-.1) node {\footnotesize $\theta1$};
\draw (-1.2,-.1) node {\footnotesize $\theta1$};
\draw (-.77,.41) node {\scriptsize $\Delta_2$};
\draw (.85,.41) node {\scriptsize $\Delta_1$};
\draw (0,1.2) node {\footnotesize $\theta0$};
\draw (0,-1.25) node {\footnotesize $\theta0$};
\end{tikzpicture}} \hspace{-1mm}\raisebox{2.2cm}{\begin{tikzpicture}[scale=.75, baseline={([yshift=.35cm]current bounding box.center)}]
\draw[thick] (-1.03,1.07) arc (240:300:2.0cm);
\draw[thick] (-1.07,1.03) arc (240:300:2.06cm);
\draw[thick] (-1.07,-1.03) arc (120:60:2.06cm);
\draw[thick] (-1.03,-1.07) arc (120:60:2cm);
\draw[thick] (0,0) circle (1.47);
\draw[thick] (0,0) circle (1.53);
\draw[fill,black] (-1.05,-1.05) circle (0.08);
\draw[fill,black] (1.05,-1.05) circle (0.08);
\draw[fill,black] (-1.05,1.05) circle (0.08);
\draw[fill,black] (1.05,1.05) circle (0.08);
\draw (1.2,0) node {\footnotesize $\theta_1$};
\draw (-1.2,0) node {\footnotesize $\theta_1$};
\draw (0,.5) node {\scriptsize $\Delta_2$};
\draw (0,-.5) node {\scriptsize $\Delta_1$};
\draw (0,1.2) node {\footnotesize $\theta_0$};
\draw (0,-1.25) node {\footnotesize $\theta_0$};
\end{tikzpicture}}\qquad\ \ 
\begin{tikzpicture}[xscale=.85,yscale=.75]
\draw (4,2.2) node {$\theta_0$};
\draw (4,-1.2) node {$\theta_0$};
\draw (5.5,0.5) node {$\theta_1$};
\draw (2.5,0.5) node {$\theta_1$};
\draw[very thick,\darkblue] (2,-2) -- (6,-2);
\draw[very thick,\darkblue] (2,3) -- (6,3);
\draw[very thick,\darkred] (2,-2) -- (2,3);
\draw[very thick,\darkred] (6,-2) -- (6,3);
\draw[very thick,\darkblue] (6,-1.6) -- (2,2.6);
\draw[very thick,\darkblue] (2,-1.6) -- (6,2.6);
\draw[very thick,->,\darkblue] (4,.5) -- (5,1.55);
\draw[very thick,->,\darkblue] (2,-1.6) -- (3,-.55);
\draw[very thick,->,\darkblue] (6,-1.6) -- (5,-.55);
\draw[very thick,->,\darkblue] (4,.5) -- (3,1.55);
\draw[thick,dashed,\darkgreen] (2,-2) -- (4.25,0.25);
\draw[thick,dashed,\darkgreen] (6,-2) -- (3.75,0.25);
\draw[thick,dashed,\darkgreen] (3.75,.75) -- (6,3);
\draw[thick,dashed,\darkgreen] (2,3) -- (4.25,0.75);
\draw (6.65,-1.5) node {\rotatebox{0}{\textcolor{\darkblue}{\footnotesize ${\phi}_1({\tau_2})$}}};
\draw (1.4,2.5) node {\rotatebox{0}{\textcolor{\darkblue}{\footnotesize ${\phi}_1({\tau_2})$}}};
\draw (1.4,-1.5) node {\rotatebox{0}{\textcolor{\darkblue}{\footnotesize  ${\phi}_2({\tau_1})$}}};
\draw (6.65,2.5) node {\rotatebox{0}{\textcolor{\darkblue}{\footnotesize  ${\phi}_1({\tau_2})$}}};
\end{tikzpicture} 
\vspace{-3mm}
\end{center}
\caption{The backreaction of two counter-propagating light-signals emitted from the north and south pole creates a gap between the future and past cosmological horizon, allowing the signals to propagate to reach the antipodal point via causal propagation. In SYK, the process is described by a four-point function with the characteristic of an OTOC. Uncrossing the chords involves quantum 6j-symbols that encode the gravitational scattering  between the two signals.}
\vspace{-0mm}
\end{figure}

What would this thought experiment look like on the SYK-side? The communication protocol is described by the following four-point function in the doubled SYK model \cite{ustwo}
\bea
\nonumber
& & \la \Psi_{\rm dS} |  {\phi}_{2} (\tau_2) \spc {\phi}_{1} (i\pi\!\spc -\!\spc \tau_2) {\phi}_{2} ({i\pi}\!\spc -\!\spc\tau_1) \spc {\phi}_{1} (\tau_1)|\Psi_{\rm dS}\ra\quad\
\eea
The initial operators ${\phi}_1(\tau_1)$ and ${\phi}_2(i\pi-\tau_1)$ create particles at the north and south pole at opposite sides of the horizon at time $\tau_1$, which are then each detected at the other side of the horizon by the corresponding operators that act at the later time $\tau_2$. The gravity description suggests that this correlator shares the characteristics of an OTOC. 

On the left, figure 6 depicts the diagrammatic representation of the OTOC in the SYK model. Following the diagrammatic rules of SYK correlators \cite{Berkooz:2018jqr}\cite{Lin:2023trc}\cite{Okuyama:2024yya}, the OTOC can be transformed back into a time-ordered correlation function by means of the quantum 6j-symbol of $U_{\!\mathsf q}(\mathfrak{sl}_2)$.  The transformation involves the product of two quantum 6j-symbols, as denoted in fig. 6 by means of the $\pm$ subscript. The 6j-symbol captures the gravitational scattering amplitude of the colliding shockwaves and equals the expectation value of a tetrahedral configuration of Wilson lines in the $SL(2,\mathbb{C})$ Chern-Simons gravity theory \cite{Witten:1989rw}. It would be a convincing check of our proposed duality to compare and match the resulting amplitude with the above semi-classical gravity description of the antipodal communication protocol.

\section*{Acknowledgments}
\vspace{-2mm}
We thank Vladimir Narovlansky for initial collaboration and Andreas Blommaert, Scott Collier, Henry Lin, Beatrix M\"uhlmann, Alex Maloney, Vladimir Narovlansky, Adel Rahman, Douglas Stanford, Lenny Susskind, Erik Verlinde, Edward Witten and Mengyang Zhang for helpful discussions. This research is supported by NSF grant PHY-2209997.

\bibliographystyle{ssg}
\addtolength{\baselineskip}{-.6mm}
\bibliography{Biblio}

\begingroup\raggedright\begin{thebibliography}{10}

\bibitem{kitaevTalks}
A.~Kitaev, ``{Talks given at the Fundamental Physics Prize Symposium and KITP
  seminars},''.
\newblock \url{https://www.youtube.com/watch?v=OQ9qN8j7EZI},
  \url{http://online.kitp.ucsb.edu/online/joint98/kitaev/},
  \url{http://online.kitp.ucsb.edu/online/entangled15/kitaev}.

\bibitem{Maldacena:2016hyu}
J.~Maldacena and D.~Stanford, ``{Remarks on the Sachdev-Ye-Kitaev model},''
  {\em Phys. Rev. D} {\bf 94} (2016), no.~10 106002,
  \href{http://xxx.lanl.gov/abs/1604.07818}{{\tt 1604.07818}}.

\bibitem{Cotler:2016fpe}
J.~S. Cotler, G.~Gur-Ari, M.~Hanada, J.~Polchinski, P.~Saad, S.~H. Shenker,
  D.~Stanford, A.~Streicher, and M.~Tezuka, ``{Black Holes and Random
  Matrices},'' {\em JHEP} {\bf 05} (2017) 118,
  \href{http://xxx.lanl.gov/abs/1611.04650}{{\tt 1611.04650}}. [Erratum: JHEP
  09, 002 (2018)].

\bibitem{Berkooz:2018jqr}
M.~Berkooz, M.~Isachenkov, V.~Narovlansky, and G.~Torrents, ``{Towards a full
  solution of the large N double-scaled SYK model},'' {\em JHEP} {\bf 03}
  (2019) 079, \href{http://xxx.lanl.gov/abs/1811.02584}{{\tt 1811.02584}}.

\bibitem{Berkooz:2018qkz}
M.~Berkooz, P.~Narayan, and J.~Simon, ``{Chord diagrams, exact correlators in
  spin glasses and black hole bulk reconstruction},'' {\em JHEP} {\bf 08}
  (2018) 192, \href{http://xxx.lanl.gov/abs/1806.04380}{{\tt 1806.04380}}.

\bibitem{HVtalks}
H.~Verlinde, ``{Talks given at the QGQC5 conference, UC Davis, August 2019, the
  Franqui Symposium, Brussels, November 2019, at `Quantum Gravity on Southern
  Cone', Argentina, December 2019, and `SYK models and Gauge Theory' workshop
  at Weizmann Institute, December 2019},''.

\bibitem{Susskind:2021esx}
L.~Susskind, ``{Entanglement and Chaos in De Sitter Space Holography: An SYK
  Example},'' {\em JHAP} {\bf 1} (2021), no.~1 1--22,
  \href{http://xxx.lanl.gov/abs/2109.14104}{{\tt 2109.14104}}.

\bibitem{Susskind:2022bia}
L.~Susskind, ``{De Sitter Space, Double-Scaled SYK, and the Separation of
  Scales in the Semiclassical Limit},''
  \href{http://xxx.lanl.gov/abs/2209.09999}{{\tt 2209.09999}}.

\bibitem{Susskind:2022dfz}
L.~Susskind, ``{Scrambling in Double-Scaled SYK and De Sitter Space},''
  \href{http://xxx.lanl.gov/abs/2205.00315}{{\tt 2205.00315}}.

\bibitem{Lin:2022nss}
H.~Lin and L.~Susskind, ``{Infinite Temperature's Not So Hot},''
  \href{http://xxx.lanl.gov/abs/2206.01083}{{\tt 2206.01083}}.

\bibitem{ustwo}
V.~Narovlansky and H.~Verlinde, ``{Double-scaled SYK and de Sitter
  Holography},'' \href{http://xxx.lanl.gov/abs/2310.16994}{{\tt 2310.16994}}.

\bibitem{Rahman:2022jsf}
A.~A. Rahman, ``{dS JT Gravity and Double-Scaled SYK},''
  \href{http://xxx.lanl.gov/abs/2209.09997}{{\tt 2209.09997}}.

\bibitem{Rahman:2023pgt}
A.~A. Rahman and L.~Susskind, ``{Comments on a Paper by Narovlansky and
  Verlinde},'' \href{http://xxx.lanl.gov/abs/2312.04097}{{\tt 2312.04097}}.

\bibitem{Rahman:2024vyg}
A.~A. Rahman and L.~Susskind, ``{Infinite Temperature is Not So Infinite: The
  Many Temperatures of de Sitter Space},''
  \href{http://xxx.lanl.gov/abs/2401.08555}{{\tt 2401.08555}}.

\bibitem{Lin:2023trc}
H.~W. Lin and D.~Stanford, ``{A symmetry algebra in double-scaled SYK},''
  \href{http://xxx.lanl.gov/abs/2307.15725}{{\tt 2307.15725}}.

\bibitem{Witten:1989ip}
E.~Witten, ``{Quantization of {Chern-Simons} Gauge Theory With Complex Gauge
  Group},'' {\em Commun. Math. Phys.} {\bf 137} (1991) 29--66.

\bibitem{Harlow:2023hjb}
D.~Harlow and T.~Numasawa, ``{Gauging spacetime inversions in quantum
  gravity},'' \href{http://xxx.lanl.gov/abs/2311.09978}{{\tt 2311.09978}}.

\bibitem{Chekhov:1999tn}
L.~Chekhov and V.~V. Fock, ``{Quantum Teichmuller space},'' {\em Theor. Math.
  Phys.} {\bf 120} (1999) 1245--1259,
  \href{http://xxx.lanl.gov/abs/math/9908165}{{\tt math/9908165}}.

\bibitem{Teschner:2003em}
J.~Teschner, ``{On the relation between quantum Liouville theory and the
  quantized Teichmuller spaces},'' {\em Int. J. Mod. Phys. A} {\bf 19S2} (2004)
  459--477, \href{http://xxx.lanl.gov/abs/hep-th/0303149}{{\tt
  hep-th/0303149}}.

\bibitem{Terashima:2011qi}
Y.~Terashima and M.~Yamazaki, ``{SL(2,R) Chern-Simons, Liouville, and Gauge
  Theory on Duality Walls},'' {\em JHEP} {\bf 08} (2011) 135,
  \href{http://xxx.lanl.gov/abs/1103.5748}{{\tt 1103.5748}}.

\bibitem{Castro:2011xb}
A.~Castro, N.~Lashkari, and A.~Maloney, ``{A de Sitter Farey Tail},'' {\em
  Phys. Rev. D} {\bf 83} (2011) 124027,
  \href{http://xxx.lanl.gov/abs/1103.4620}{{\tt 1103.4620}}.

\bibitem{goldman1984symplectic}
W.~M. Goldman, ``The symplectic nature of fundamental groups of surfaces,''
  {\em Advances in Mathematics} {\bf 54} (1984), no.~2 200--225.

\bibitem{Witten:1988hc}
E.~Witten, ``{(2+1)-Dimensional Gravity as an Exactly Soluble System},'' {\em
  Nucl. Phys. B} {\bf 311} (1988) 46.

\bibitem{kauffman1988new}
L.~H. Kauffman, ``New invariants in the theory of knots,'' {\em The American
  mathematical monthly} {\bf 95} (1988), no.~3 195--242.

\bibitem{reshetikhin1991invariants}
N.~Reshetikhin and V.~G. Turaev, ``Invariants of 3-manifolds via link
  polynomials and quantum groups,'' {\em Inventiones mathematicae} {\bf 103}
  (1991), no.~1 547--597.

\bibitem{Witten:1989rw}
E.~Witten, ``{Gauge Theories, Vertex Models and Quantum Groups},'' {\em Nucl.
  Phys. B} {\bf 330} (1990) 285--346.

\bibitem{Gaiotto:2014lma}
D.~Gaiotto, ``{Open Verlinde line operators},''
  \href{http://xxx.lanl.gov/abs/1404.0332}{{\tt 1404.0332}}.

\bibitem{roger2014skein}
J.~Roger and T.~Yang, ``The skein algebra of arcs and links and the decorated
  Teichm{\"u}ller space,'' {\em Journal of Differential Geometry} {\bf 96}
  (2014), no.~1 95--140.

\bibitem{ustwotwo}
H.~Verlinde and M.~Zhang, ``SYK Correlators from de Sitter Gravity,''
  \href{http://xxx.lanl.gov/abs/to appear}{{\tt to appear}}.

\bibitem{penner1987decorated}
R.~C. Penner, ``The decorated Teichm{\"u}ller space of punctured surfaces,''
  {\em Communications in Mathematical Physics} {\bf 113} (1987) 299--339.

\bibitem{Blommaert:2023opb}
A.~Blommaert, T.~G. Mertens, and S.~Yao, ``{Dynamical actions and
  q-representation theory for double-scaled SYK},''
  \href{http://xxx.lanl.gov/abs/2306.00941}{{\tt 2306.00941}}.

\bibitem{tHooft:1996ziz}
G.~'t~Hooft, ``{Quantization of point particles in (2+1)-dimensional gravity
  and space-time discreteness},'' {\em Class. Quant. Grav.} {\bf 13} (1996)
  1023--1040, \href{http://xxx.lanl.gov/abs/gr-qc/9601014}{{\tt
  gr-qc/9601014}}.

\bibitem{Harlow:2018tqv}
D.~Harlow and D.~Jafferis, ``{The Factorization Problem in Jackiw-Teitelboim
  Gravity},'' {\em JHEP} {\bf 02} (2020) 177,
  \href{http://xxx.lanl.gov/abs/1804.01081}{{\tt 1804.01081}}.

\bibitem{Hikida:2021ese}
Y.~Hikida, T.~Nishioka, T.~Takayanagi, and Y.~Taki, ``{Holography in de Sitter
  Space via Chern-Simons Gauge Theory},'' {\em Phys. Rev. Lett.} {\bf 129}
  (2022), no.~4 041601, \href{http://xxx.lanl.gov/abs/2110.03197}{{\tt
  2110.03197}}.

\bibitem{Hikida:2022ltr}
Y.~Hikida, T.~Nishioka, T.~Takayanagi, and Y.~Taki, ``{CFT duals of
  three-dimensional de Sitter gravity},'' {\em JHEP} {\bf 05} (2022) 129,
  \href{http://xxx.lanl.gov/abs/2203.02852}{{\tt 2203.02852}}.

\bibitem{Okuyama:2024yya}
K.~Okuyama, ``{Doubled Hilbert space in double-scaled SYK},''
  \href{http://xxx.lanl.gov/abs/2401.07403}{{\tt 2401.07403}}.

\bibitem{Verlinde:1989ua}
H.~L. Verlinde, ``{Conformal Field Theory, 2-$D$ Quantum Gravity and
  Quantization of Teichmuller Space},'' {\em Nucl. Phys. B} {\bf 337} (1990)
  652--680.

\bibitem{Klemm:2002ir}
D.~Klemm and L.~Vanzo, ``{De Sitter gravity and Liouville theory},'' {\em JHEP}
  {\bf 04} (2002) 030, \href{http://xxx.lanl.gov/abs/hep-th/0203268}{{\tt
  hep-th/0203268}}.

\bibitem{Collier:2023fwi}
S.~Collier, L.~Eberhardt, and M.~Zhang, ``{Solving 3d Gravity with Virasoro
  TQFT},'' {\em SciPost Phys.} {\bf 15} (2023) 151,
  \href{http://xxx.lanl.gov/abs/2304.13650}{{\tt 2304.13650}}.

\bibitem{Collier:2023cyw}
S.~Collier, L.~Eberhardt, B.~M\"uhlmann, and V.~A. Rodriguez, ``{The Virasoro
  Minimal String},'' \href{http://xxx.lanl.gov/abs/2309.10846}{{\tt
  2309.10846}}.

\bibitem{spacelike-lcft-string}
S.~Collier, L.~Eberhardt, B.~M\"uhlmann, and V.~A. Rodriguez, ``{to appear},''.

\bibitem{douglas-talk-kitp}
D.~Stanford, ``{Talk given at KITP, 2018},''.
\newblock
  \url{https://online.kitp.ucsb.edu/online/chord18/doublescale/rm/jwvideo.html}.

\bibitem{Mertens:2017mtv}
T.~G. Mertens, G.~J. Turiaci, and H.~L. Verlinde, ``{Solving the Schwarzian via
  the Conformal Bootstrap},'' {\em JHEP} {\bf 08} (2017) 136,
  \href{http://xxx.lanl.gov/abs/1705.08408}{{\tt 1705.08408}}.

\bibitem{Lam:2018pvp}
H.~T. Lam, T.~G. Mertens, G.~J. Turiaci, and H.~Verlinde, ``{Shockwave S-matrix
  from Schwarzian Quantum Mechanics},'' {\em JHEP} {\bf 11} (2018) 182,
  \href{http://xxx.lanl.gov/abs/1804.09834}{{\tt 1804.09834}}.

\bibitem{Hotta:1992qy}
M.~Hotta and M.~Tanaka, ``{Shock wave geometry with nonvanishing cosmological
  constant},'' {\em Class. Quant. Grav.} {\bf 10} (1993) 307--314.

\bibitem{Leblond:2002ns}
F.~Leblond, D.~Marolf, and R.~C. Myers, ``{Tall tales from de Sitter space 1:
  Renormalization group flows},'' {\em JHEP} {\bf 06} (2002) 052,
  \href{http://xxx.lanl.gov/abs/hep-th/0202094}{{\tt hep-th/0202094}}.

\bibitem{Aalsma:2021kle}
L.~Aalsma, A.~Cole, E.~Morvan, J.~P. van~der Schaar, and G.~Shiu, ``{Shocks and
  information exchange in de Sitter space},'' {\em JHEP} {\bf 10} (2021) 104,
  \href{http://xxx.lanl.gov/abs/2105.12737}{{\tt 2105.12737}}.

\bibitem{Anegawa:2023dad}
T.~Anegawa and N.~Iizuka, ``{Shock waves and delay of hyperfast growth in de
  Sitter complexity},'' {\em JHEP} {\bf 08} (2023) 115,
  \href{http://xxx.lanl.gov/abs/2304.14620}{{\tt 2304.14620}}.

\bibitem{Gao:2016bin}
P.~Gao, D.~L. Jafferis, and A.~C. Wall, ``{Traversable Wormholes via a Double
  Trace Deformation},'' {\em JHEP} {\bf 12} (2017) 151,
  \href{http://xxx.lanl.gov/abs/1608.05687}{{\tt 1608.05687}}.

\end{thebibliography}\endgroup

\end{document}